\documentclass[fleqn,usenatbib]{mnras}

\usepackage{newtxtext,newtxmath}

\usepackage[T1]{fontenc}

\DeclareRobustCommand{\VAN}[3]{#2}
\let\VANthebibliography\thebibliography
\def\thebibliography{\DeclareRobustCommand{\VAN}[3]{##3}\VANthebibliography}

\usepackage{graphicx}	
\usepackage{amsmath}	
\usepackage{caption}
\usepackage{booktabs,threeparttable}

\newcommand{\zfourge}{{\tt ZFOURGE}}
\newcommand{\zfire}{{\tt ZFIRE}}
\newcommand{\illustris}{{\tt IllustrisTNG}}

\newcommand{\logMstarMsun}{{$\log({\rm M}_{\star}/{\rm M}_{\odot})$}}
\newcommand{\logMgasin}{$\log \dot{\rm{M}}_{\rm{in, gas}}$}
\newcommand{\inflow}{$\rm{M_\odot Gyr^{-1}}$}
\graphicspath{{./}{figures/}}

\title[Gas Inflow History in Cluster and Field Halos]{\zfire\, - The Gas Inflow Inequality for Satellite Galaxies in Cluster and Field Halos at $z=2$}

\author[A. Harshan et al.]{
Anishya Harshan,$^{1,2}$
Kim-Vy Tran,$^{1,2}$
Anshu Gupta$^{2,3}$
Glenn G. Kacprzak$^{2,4}$
Themiya Nanayakkara$^{2,4}$
\\
$^{1}$School of Physics, University of New South Wales, Sydney, NSW 2052, Australia\\
$^{2}$ARC Centre of Excellence for All Sky Astrophysics in 3 Dimensions (ASTRO 3D), Australia\\
$^{3}$International Centre for Radio Astronomy Research (ICRAR), Curtin University, Bentley WA, Australia\\
$^{4}$Centre for Astrophysics and Supercomputing, Swinburne University of Technology, Hawthorn, VIC 3122, Australia\\
}

\date{Accepted XXX. Received YYY; in original form ZZZ}

\pubyear{2022}

\begin{document}
\label{firstpage}
\pagerange{\pageref{firstpage}--\pageref{lastpage}}
\maketitle

\begin{abstract}
Gas inflow into galaxies should affect the star formation and hence the evolution of galaxies across cosmic time. In this work, we use TNG100 of the \illustris\ simulations to understand the role of environment on gas inflow rates in massive galaxies at $z\geq2$. We divide our galaxies (\logMstarMsun $\geq 10.5$) into cluster ($\log \rm{M_{halo}/M_\odot}\geq 13$) and field ($\log \rm{M_{halo}/M_\odot}<13$) galaxies at $z=2$ and further divide into centrals and satellites. We track their gas inflow rates from $z=6$ to 2 and find that the total gas inflow rates of satellite galaxies rapidly decline after their infall into cluster halos and as they reach the cluster center. At $z=2$, the gas inflow rate of cluster satellite galaxies is correlated with the cluster-centric radii and not the host halo mass. In contrast, the gas inflow rate in centrals is strongly correlated with the host halo mass at $z\geq2$. Our study indicates that between redshifts 6 to 2, the gas that normally is accreted by the satellite galaxies is redirected to the center of the cluster halo as inflows to the cluster centrals and forming the intra-cluster medium.  Our analysis suggest that the inequality of gas accretion between massive satellite and central galaxies is responsible for the starvation of cluster satellite galaxies that evolve into the massive quenched cluster galaxies observed at $z<0.5$.
\end{abstract}




\section{Introduction}

Understanding gas flows is quintessential to understanding the evolution of star formation in a galaxy and hence galaxy growth. In the $\Lambda$CDM model of the universe, pristine or metal-poor gas is accreted onto galaxies as fuel for star formation over cosmic time. As star formation progresses, the gas reservoir of the galaxy is depleted and energetically ejected in form of winds and outflows driven by star formation or the active-galactic nuclei (AGN). The ejected gas is metal-enriched and may either be accreted back onto the galaxy or enrich the circumgalactic medium (CGM) and the inter-galactic medium (IGM). The progression of the baryonic cycle across cosmic time describes the properties of galaxies, and yet remains poorly understood.

Numerical simulations predict that gas accretion in a galaxy can proceed in several ways. Gas accreted onto a massive galaxy ($\log M_{h}/M_\odot >11$) from the surrounding IGM is shock heated and will eventually radiatively cool \citep{Rees1977,Silk1977} and settle onto the galactic disk. This is called ``hot-mode" accretion. Alternatively, pristine gas can flow into the galaxy via filaments from the cosmic web and is called ``cold-mode" accretion  \citep{Keres2005, Keres2009, Nelson2013, Stern2020, Wright2021}. Galaxies can also acquire gas from mergers with other galaxies \citep{Angles2017, Martin2022}. The different modes of gas accretion occur simultaneously and leave different imprints on the evolution of galaxies and their star formation .


A series of paper from the \zfire\ survey suggests that, the environment of a galaxy has been shown to correlate with star formation rates and interstellar medium properties of galaxies as early as $z=2$ \citep{Tran2010,  Kawinwanichakij2017b, Forrest2017a,Alcorn2019a, Watson2019,Harshan2020j, Harshan2021a}. Gravitational and hydrodynamical interactions with other galaxies and the intra-cluster medium (ICM) in dense environments preferentially affects the gas in satellite galaxies \citep{Gunn1972, Moore1996,Fumagalli2009, Cortese2021}. At lower redshifts $z<0.8$, studies find a deficiency of atomic and molecular gas fractions in galaxies in high density environments  \citep{ Scott2013, Peng2015, Brown2017a}. In contrast, studies of galaxies in proto-clusters at $z>1.5$ find higher gas fractions \citep{Noble2017a, Rudnick2017a, Hayashi2017,Aoyama2022}. However, measuring gas fractions in galaxies at $z>1$ remains challenging due to sensitivity limits and access to key spectral lines.

Environment plays a crucial role in regulating the gas accretion history of individual galaxies. Numerical simulations suggest that galaxies residing in the filaments of the cosmic web acquire their gas mass through cold-mode accretion \citep{Birnboim2003, Ocvirk2008, Keres2009,Liao2019}. Using cosmological simulations, studies like \cite{Simha2009} and \cite{VandeVoort2017} find that satellite galaxies in massive halos have lower gas accretion compared to central galaxies. \cite{VandeVoort2017} also find that the gas inflow rate in central galaxies is not strongly affected by the environment of the galaxy. In a recent study, \cite{Chun2020} show that at $z<5$, when the gas in the cosmic web is ionized, accretion of gas is a function of surrounding gas density and the position/proximity of the galaxy to the cosmic web. On the other hand, galaxies that fall into massive halos lose their gas reservoir due to interaction with the cluster potential, the ICM, or nearby galaxies.  

Observational studies of gas inflows remain challenging at higher redshifts and have relied on the detection of Ly$\alpha$ emission around quasars and protoclusters, or quasar sightlines \citep{Dijkstra2009, Bouche2013,  Kacprzak2016a, Zabl2019, Umehata2019, Fu2021, Daddi2021, Daddi2022}. Additionally, interpretation of the observational results of inflows is made complicated by the effect of outflows, active galactic nuclei (AGN), and star formation. Studies using Ly$\alpha$ emission line find a higher occurrence of Ly$\alpha$ emitting halos around high density environments at $z>2$ indicating active gas accretion \citep{Umehata2019,  Daddi2021, Daddi2022}, however, accretion in isolated galaxies in the high redshift universe is not easily observable.  

In \cite{Harshan2021a}, using \illustris\ we show that the massive galaxies in the proto-cluster environment at $z\sim2$ have been gas poor since $z\sim 5$ compared to their field counterparts. We hypothesize that lower gas fractions in massive proto-cluster galaxies directly result in lower star formation in cluster environment as measured using the star formation histories from \zfourge\ - \zfire\ observations \citep{Straatman2016a, Nanayakkara2016, Tran2015}. The lower gas fraction can arise due to a lower inflow of gas from the surrounding medium either due to feedback, \citep{Zinger2020} or the position of the galaxy in the cosmic web, gas poor mergers, and outflows. In this work, we test our hypothesis by tracking how gas inflow onto individual galaxies varies with the environment.


This paper is organized as follows. In section \ref{sec:methodology}, we describe the data from cosmological simulations \illustris\ and methodology to describe gas inflows. In sections \ref{sec:results} and \ref{sec:discussion} we state our results and discussion, and in section \ref{sec:summary} summarise the results. 

\section{IllustrisTNG and Methodology}
\label{sec:methodology}


\illustris\ is a suite of magneto-hydro-dynamical cosmological simulations based on the $\Lambda$CDM cosmology \citep{Pillepich2018b, Nelson2018, Springel2018a,  Marinacci2017, Naiman2017}. \illustris\ extends the Illustris framework with  kinetic  black  hole  feedback,  magneto-hydrodynamics, and a revised scheme for galactic winds, among other changes \citep{Weinberger2017, Pillepich2018}. It spans a range of galaxy environments including halo masses as high as $\rm{M}_{halo} \sim 10^{14.6} \rm{M}_\odot$ at $z=0$.

In this work, we use TNG100 of the \illustris project with $\rm{L}_{box} = 110.7$ cMpc, and a total volume $\sim10^6 ~\rm{Mpc}^3 $, to map the gas inflow rate histories of the cluster and field galaxies at $z=2$. The data of TNG100 is publicly available and described by \cite{Nelson2019a}. The TNG100 simulation has a baryonic mass resolution of $\rm{m}_b = 1.4\times 10^6 ~\rm{M}_\odot$ which provides about 1000 stellar particles per galaxy for a galaxy with stellar mass \logMstarMsun $= 9$, and proportionally more stellar particles for more massive galaxies. 

Following the selection in \cite{Harshan2021a}, we select galaxies associated with halos of $\rm{M}_{halo} \geq 10^{13} \rm{M}_\odot$ as cluster galaxies and galaxies in halos $\rm{M}_{halo} < 10^{13} \rm{M}_\odot$ as field galaxies at $z=2$. 

In observational studies, field galaxies are defined as galaxies in relatively under-dense environments including smaller group halos. We further divide the cluster and field sample into central and satellite populations. The central galaxy of each halo is defined as the most massive galaxy in the friends-of-friends (FOF) group (halo) and satellite galaxies in each halo are all galaxies except the central in the FOF group (halo). 

In \cite{Harshan2021a}, we find that only the most massive galaxies show a consistently low total gas fraction in the progenitors and hence, for this analysis we are only considering massive galaxies at $\log \rm{M}_*/\rm{M}_\odot \geq 10.5$ in both cluster and field halos. To separate the effect of stellar mass and environment, we have mass matched the field sample (both centrals and satellites) with the cluster satellite sample (Table \ref{table:sample}). Mass matching is done by selecting all galaxies (not one-to-one) in the field sub-samples within $\pm0.01$ dex range of each cluster satellite galaxy. We excluded cluster satellites without a stellar mass match in the field samples. Ultimately, we obtain a sample of 394 galaxies in the field centrals, 69 field satellites and 44 cluster satellite galaxies. Due to the stellar-to-halo mass relation \citep{Moster2018, Behroozi2019, Pillepich2018b}, central galaxies of massive clusters are hosted by more massive halos than central galaxies in the field. Thus the central galaxies in cluster halos are significantly more massive than the rest of the sample, hence we are unable to mass match the cluster central sample to the cluster satellite sample meaningfully (Sample sets are summarised in Table \ref{table:sample}).

\begin{figure}
    \begin{center}

    \includegraphics[scale = 0.5]{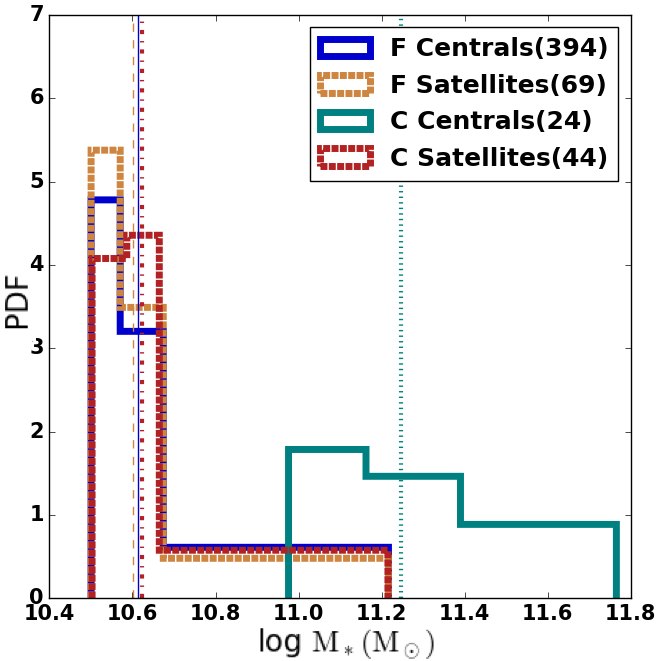}
    
    \end{center}
    
    \caption{The stellar mass distribution of field ($\log \rm{M_{halo}/M_\odot}<13$) centrals (solid blue line) and field satellites (dashed yellow line) stellar mass matched with the cluster ($\log \rm{M_{halo}/M_\odot}\geq13$) satellites (red dashed line) from TNG100 of \illustris\ at $z=2$. The solid teal line shows the stellar mass distribution of the cluster centrals. The dashed vertical lines indicate the median value of each sample indicated by the respective color and the number of galaxies in each sample is indicated in the label. The histograms are normalised by the total number of counts and the bin width. }
    \label{fig:smd}
\end{figure}

The stellar mass distribution of four samples is presented in Figure \ref{fig:smd} with median stellar mass for each sample indicated with vertical dashed lines. The median stellar mass for field centrals (\logMstarMsun $= 10.6 \pm 0.1$), field satellites (\logMstarMsun $= 10.6 \pm 0.1$) and cluster satellites (\logMstarMsun $= 10.6 \pm 0.1$) fall within $1\sigma$ deviation. The median stellar mass for the cluster centrals is \logMstarMsun $= 11.2 \pm 0.2$.  We perform the Kolmogorov-Smirnov test to test the equality of the mass matched sample (excluding the massive cluster central population).  We calculate the K-S test statistic of $0.1$ and p-value = $0.73$  between the field centrals and cluster satellites and 0.15 (K-S statistic) and 0.50 (p-value) between the field and cluster satellites. Thus the stellar mass distribution of selected field centrals and satellites conforms to the cluster satellites. Throughout the paper, field centrals are referred to as `F centrals', field satellites as `F satellites', cluster centrals are referred to as `C centrals' and cluster satellites as `C satellites' in the figure legends. 

\begin{table*}

\begin{threeparttable}
    \label{table:sample}
    \begin{tabular}{lccccr} 
        \hline
	Sample & Sample Size & \logMstarMsun\tnote{a} & Gas Radii\tnote{b} & $ \rm{R/R}_{200}$\tnote{c} & $\log D_5$\tnote{d}\\
        \hline
        \hline
        F Satellites  & 69 & $ 10.6 \pm 0.1$ & $15.8 \pm 0.8$ & $0.9 \pm0.1$ & $2.0\pm 0.1$  \\
        F Centrals & 394 & $ 10.6 \pm 0.1$ & $24.5 \pm 1.0$  & - & $2.2\pm0.0$ \\
        C Satellites & 44 & $ 10.6 \pm 0.1$ & $17.3 \pm 1.3$ & $0.5 \pm 0.2$ & $1.9\pm0.0$\\
        C Centrals \tnote{e} & 24& $ 11.2 \pm 0.2$  & $27.6 \pm 2.4 $ & - & $1.6\pm0.1$\\
    
	\hline
    \end{tabular}

\caption{Summary of selected centrals and satellites sample in field and cluster environments}

    \begin{tablenotes}
        \item[a] Median log Stellar Mass of the sample
        \item[b] Median Gas Radii is presented in kpc
        \item[c] Median Halo-centric Distance (R) normalised by the $R_{200}$ of the host halo
        \item[d] Median log of fifth nearest neighbour distance is presented in kpc
        \item[e] C Centrals are not stellar mass matched with C Satellites
    \end{tablenotes}
\end{threeparttable}

\end{table*}

\subsection{Measuring Gas Inflows}
\label{sec:inflowcalc}

We measure the gas radii of the galaxies by calculating the gas density profile of each galaxy within 10 times the stellar half mass radii ($10 R_{h}$), in order to minimize the contribution of gas from the halo itself. We consider the radii where the gas density first drops to within 10\%  of the average gas density at $10 R_{h}$ as the gas radii of the galaxy. Although the satellite galaxies show a clear flattening of the gas density profiles, central galaxies do not. The gas in the halo is associated to the central galaxy which would describe gas density profile and the higher gas radii in the centrals compared to the stellar mass matched satellites. The median gas radii of the four samples are reported in Table \ref{table:sample}. The median gas radii for the entire stellar mass matched sample (i.e., field centrals, field satellites and cluster satellites) is $22.4\pm 0.8$ kpc. 

Inside the gas disk, the gas turbulence is mainly driven by radial flows. Whereas, gas accretion affects the outskirts of the gas disk \citep{Forbes2022}. Thus, we calculate the inflowing gas at approximately the gas radii. We calculate the inflowing gas at 20 kpc radii in a 5 kpc shell by calculating the radial velocity of each gas cell in the prescribed shell and measure the total gas mass inflow rate at a specific redshift as:
\begin{equation} 
\dot{\rm{M}}_{\rm{in, gas}}(\rm{z}) = \sum_{i=0}^{N}{\rm{v_{rad}}_{i,z}\rm{M}_{i,z}/\Delta r }
\end{equation}

where $\rm{v_{rad}}_{i,z}$ is the radial velocity of each gas cell of mass $\rm{M}_{i,z}$ at radius $20<\rm{r_i}<25$ kpc from the center of the galaxy at each redshift $z$. The radial velocity is calculated with respect to the center of the galaxy taking into account the motion of the galaxy and the local Hubble expansion. Additionally, we measure the gas inflow rates using gas cells with conservative threshold of $\rm{v_{rad}}<-50 \rm{km/s}$ in the considered shell. 

\begin{figure*}
    \begin{center}

    \includegraphics[scale = 0.26]{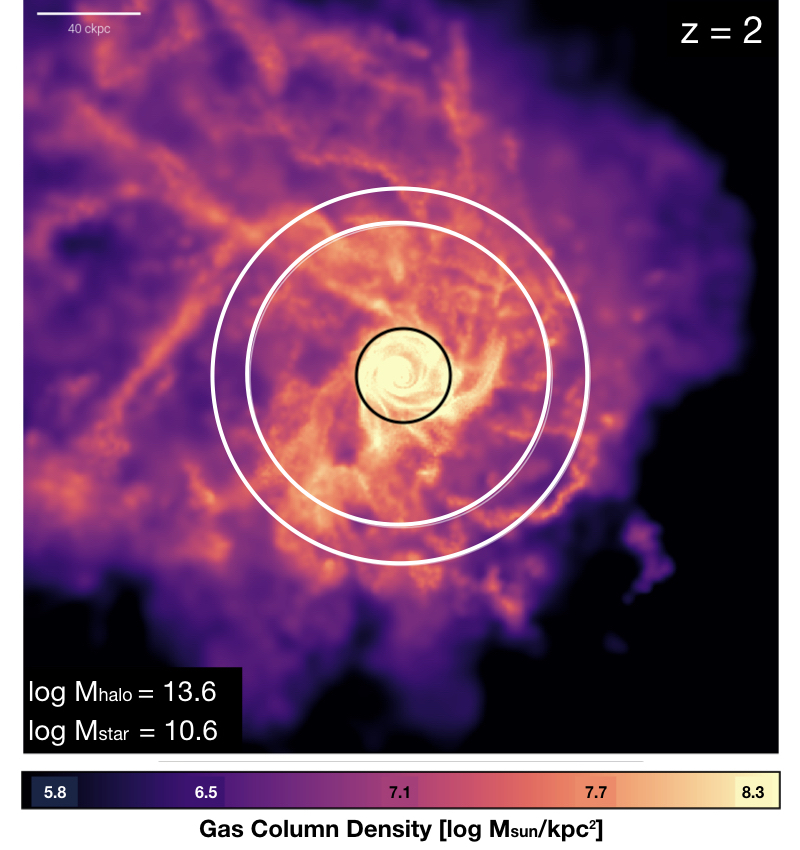}
    \includegraphics[scale = 0.26]{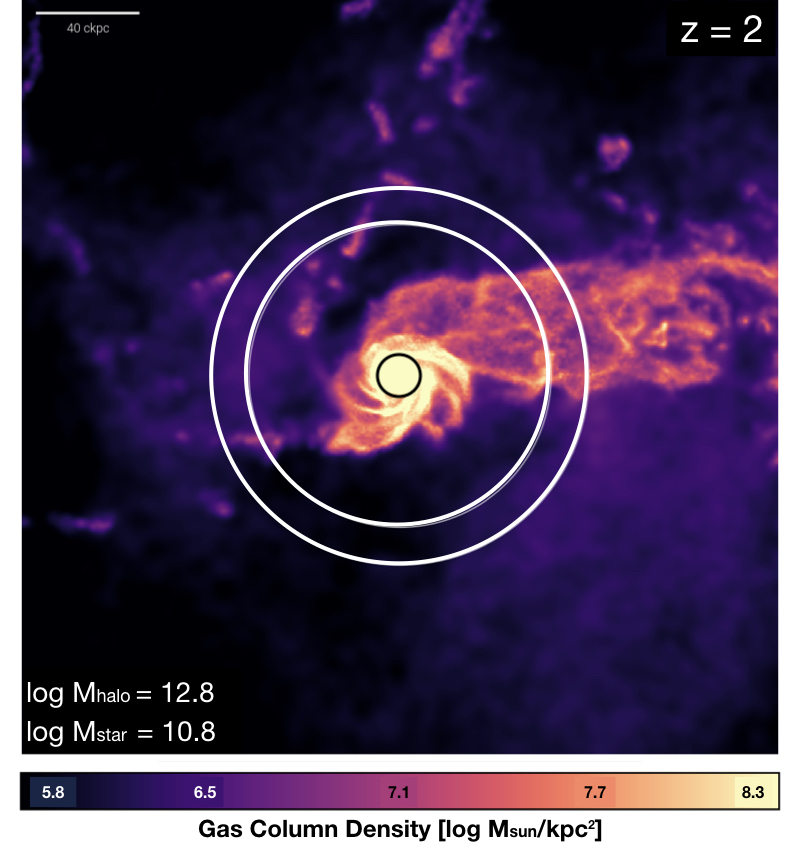}
    \includegraphics[scale = 0.26]{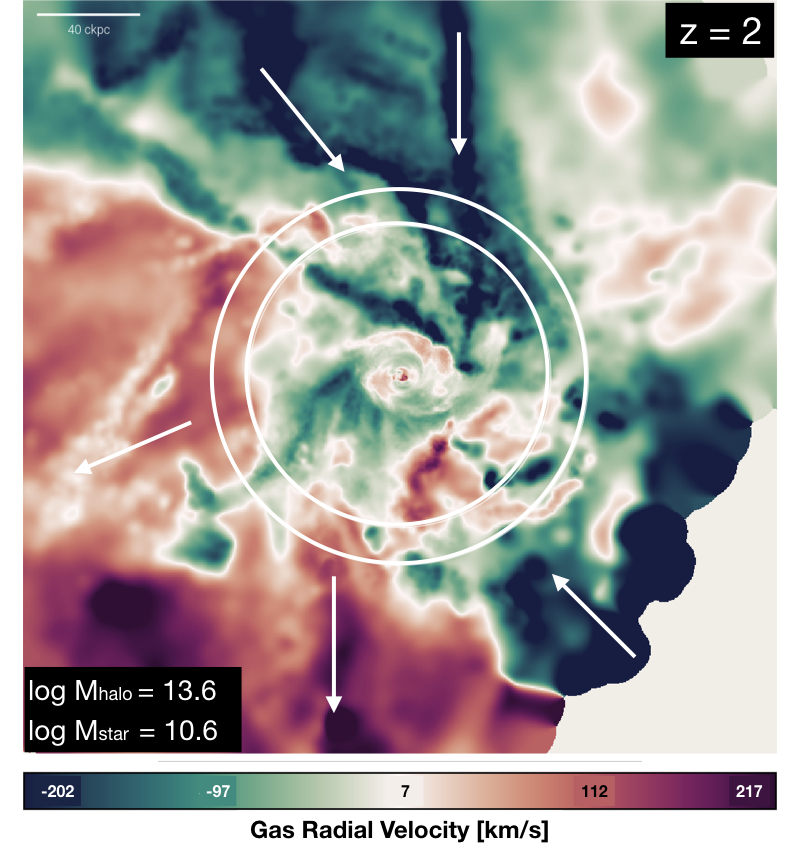}
    \includegraphics[scale = 0.26]{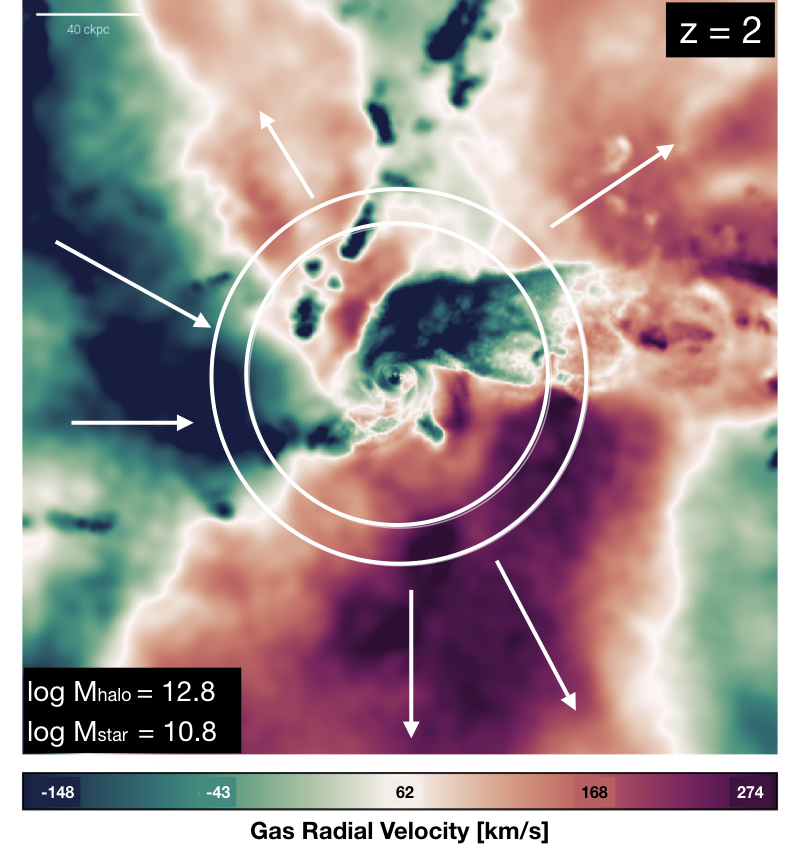}
    \end{center}
    \caption{Gas column density (top) and radial velocity (bottom) distribution of gas in two example galaxies at z = 2 (left: satellite; right: central) projected face-on. The black circles (top panels) show twice the stellar half mass radius of the galaxy and white shells depict the 20 kpc and 25 kpc radius from the center of the galaxy. We calculate the rate of inflowing gas in the 5 kpc shell at 20 kpc radii from the center of the galaxy. The inflowing gas is shown in green with darkness increasing with the increasing radial velocity and the outflowing gas is depicted in red. The images are 100 kpc x 100 kpc at $z=2$ created using the \illustris\ data visualisation tool (https://www.tng-project.org/data/vis/; credit: TNG Collaboration).}
    \label{fig:gasex}
\end{figure*}

Figure \ref{fig:gasex} shows an example of a central and a satellite galaxy. The top panel shows the column density of total gas with density increasing from purple to yellow. The bottom panel show the radial velocity of the gas cells associated with the galaxy such that the inflowing gas is colored in green and the outflowing gas is red. The images are overlayed with the spherical shell shown in white in which the gas inflow rate is calculated.

We use data from the SUBLINK algorithm which tracks the merger histories of the galaxies \citep{Rodriguez-Gomez2015} to quantitatively follow the evolution of galaxies and consider only the main progenitor branch (considering the most massive progenitors at each snapshot) to calculate the gas inflow rates and associated histories of the galaxies.

\section{Results: History of Gas Inflow Rates}
\label{sec:results}

\begin{figure}
    \begin{flushleft}
    \includegraphics[scale = 0.52]{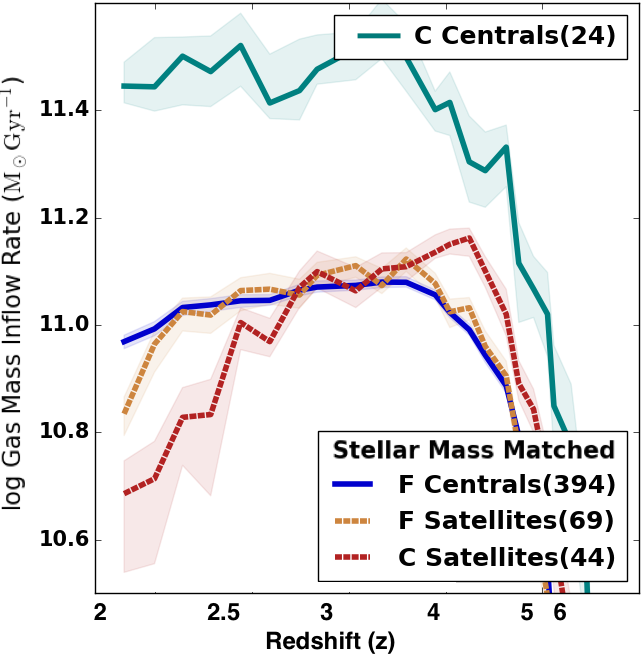}
    \end{flushleft}
    
    \caption{ Gas inflow rate histories of massive galaxies selected at $z=2$ depend on their position in the halo. The solid lines show the bootstrapped median of gas inflow histories and the shaded region shows the $1\sigma$ error. The gas inflow rates of satellite galaxies in cluster halos (red) are higher at $z>4$ compared to the field centrals (blue). The rate of gas inflow of total gas decreases rapidly in the satellite galaxies associated with larger cluster halos at $z\gtrsim3$.  The number of galaxies in each bin is given in parenthesis in each label.}
    \label{fig:gas_in}
\end{figure}

We measure the history of total gas inflow rates of satellite and central galaxies in field ($\rm{M}_{halo} < 10^{13} \rm{M}_\odot$) and cluster ($\rm{M}_{halo} \geq 10^{13} \rm{M}_\odot$) halos at  $z \geq 2 $ as described in section \ref{sec:inflowcalc} and show in Figure \ref{fig:gas_in}. The stellar mass matched field satellites (Yellow dashed line;  Figure \ref{fig:gas_in}) and field centrals (Blue solid line ;  Figure \ref{fig:gas_in}) have a similar history of gas inflow rates. In contrast, the stellar mass matched cluster satellite galaxies (Red dashed line;  Figure \ref{fig:gas_in}) have higher gas inflow rates compared to the field (centrals + satellites) sample at $z>4$, after which the gas inflow rates in cluster satellites decreases ($\approx 0.5$ dex decline) by $z=2$. The decline in gas inflow rates in cluster satellites indicates an effect of environment in the early universe (at $6>z>2$). However, this effect is measured in satellites of cluster halos only. 

Cluster centrals (Teal solid line;  Figure \ref{fig:gas_in}) have an increasing gas inflow rate at $6>z>3$, after which the gas inflow rate is constant for the next 1.3 Gyrs. By $z=2$, cluster centrals have an $\approx 0.5$ dex higher gas inflow rate compared to the field centrals (Table \ref{table:inflow}). However, cluster centrals are not stellar mass matched with the other three sample sets, therefore it is difficult to disentangle if their stellar mass and/or environment is driving their higher gas mass inflow rate. Within the stellar mass matched sample, in the appendix \ref{appendix} we explore the effect of stellar mass of the galaxy on the gas inflow rates. We find that at $z\leq3$, the stellar mass of the galaxy has no significant correlation with the gas inflow rates before and after infall.

\begin{table*}
\begin{threeparttable}
    
    \label{table:inflow}
    \begin{tabular}{lllll}
    \hline
    Sample &   $z=2$ & $z=3$&  $z=4$ &$z=6$ \\ 
    \hline \hline
    F Satellites  & $10.8\pm 0.1$ & $11.1 \pm 0.2$ &  $11.0\pm0.2$ & $10.2\pm0.6$\\
    F Centrals    & $10.9\pm 0.01$ & $11.1 \pm 0.2$ &  $11.0\pm0.3$ & $10.2\pm0.6$\\
    C Satellites  & $10.7\pm 0.7$ & $11.1 \pm 0.3$ &  $11.2\pm0.2$ & $10.4\pm0.6$\\
    C Centrals    & $11.4\pm 0.3$ & $11.5 \pm 0.3$ &  $11.4\pm0.5$ & $10.8\pm0.6$\\
    
    \hline
    \end{tabular}
    
    \caption{Log of gas inflow rates (\inflow) at z = 2, 3, 4 and 6}

\end{threeparttable}   
\end{table*}

\subsection{Gas Inflow Rates: Before and After Infall}
\label{sec:bi_ai}

In Figure \ref{fig:gas_in}, we find that the cluster satellites undergo a drastic change and the gas inflow rates decrease by $\approx 0.5$ dex in $\sim1.5$ Gyrs (at redshifts 4 to 2), but the same is not observed in field satellites. In this section, we investigate how the gas inflow rates of satellite galaxies in cluster ($\rm{M}_{halo} \geq 10^{13} \rm{M}_\odot$) and field ($\rm{M}_{halo} < 10^{13} \rm{M}_\odot$) halos change before and after infall into their host halo at $z=2$. 

\begin{figure*}
    \begin{center}
    \includegraphics[width = \textwidth]{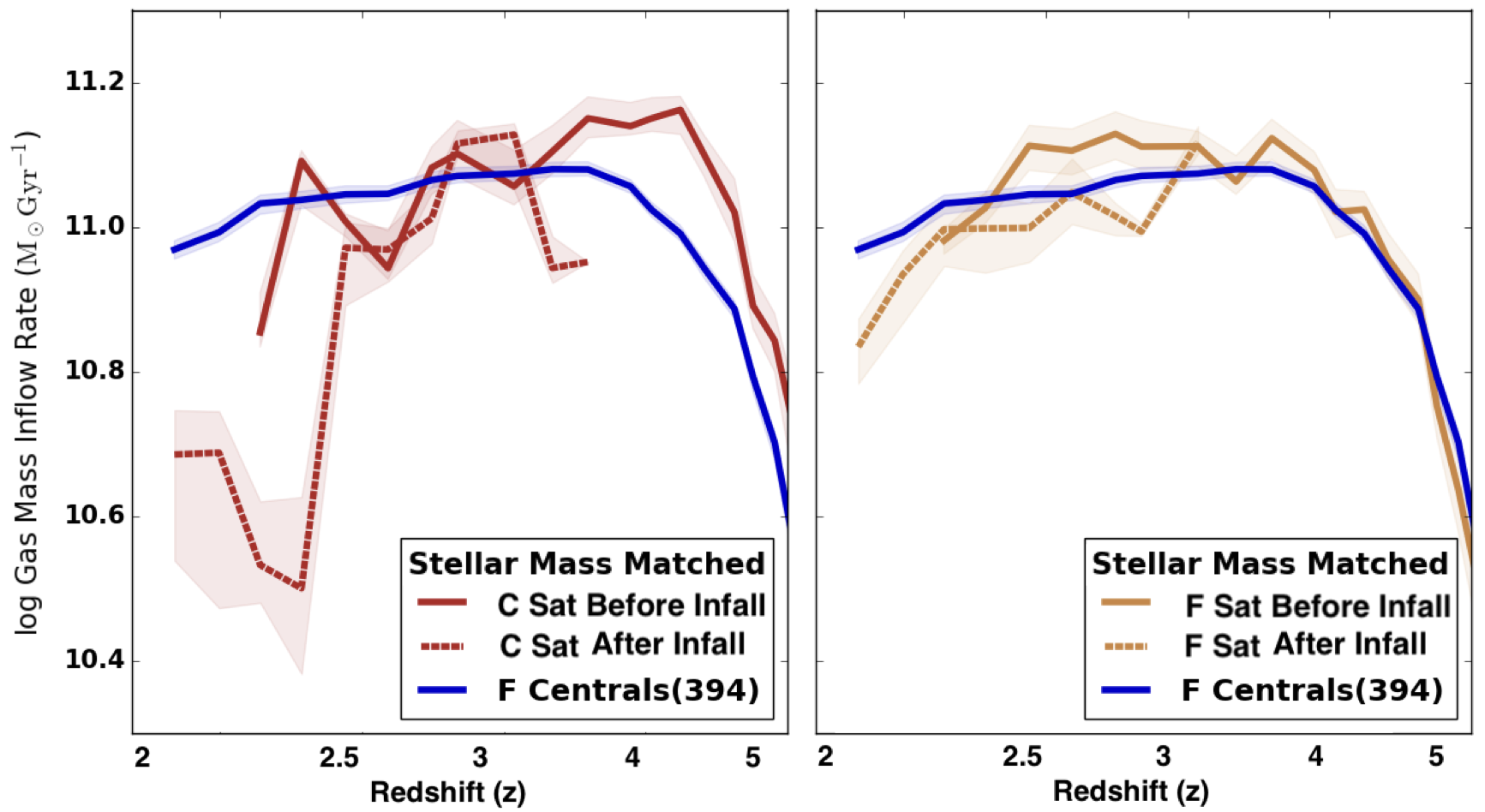}
   
    \end{center}
    
    \caption{ The gas inflow rates in cluster satellites rapidly decline after infall, in contrast, the stellar mass matched field satellites have comparable history of gas inflow rates with the field centrals before and after infall. The red (left) and yellow (right) lines show the history of gas inflow rates of cluster (left) and field (right) satellites before (solid) and after (dashed) infall into the host halo at $z=2$ respectively. The solid blue lines in both panels show the history of gas inflow rates for field centrals. Shaded regions show $1\sigma$ of the distribution.   }
    \label{fig:infall_inflow}
\end{figure*}

We measure the infall time of satellite galaxies at $z=2$ in the following manner. We assume that the merger tree of the most massive galaxy in a FOF halo also tracks the merger history of the FOF halo \citep{Gupta2018}. We reconstruct the main progenitor branch history by identifying the host halo of the progenitor of the most massive galaxy at each previous snapshot between $2\leq z \leq6$. We calculate the infall times of each satellite galaxy as the snapshot when the galaxy becomes bound to the progenitor halo. Figure \ref{fig:infall_inflow} shows the gas inflow rates of the cluster satellites (left panel) and field satellites (right panel) before and after their infall into the host halo at $z=2$.

Before infall into the host halo at $z=2$, cluster satellite sample have higher gas inflow rates compared to the stellar mass matched field centrals at $z\gtrsim3$. At $z\lesssim3$, the gas inflow rates of cluster satellites sample closely follows that of the field centrals. However, Gas inflow rates of field satellites closely follow that of the field centrals throughout their history between $6>z>2$ (yellow solid line; Figure \ref{fig:infall_inflow} right panel). 

After infall into the host halo at $z=2$, cluster satellites (red dashed line; Figure \ref{fig:infall_inflow} left panel) have rapidly declining gas inflow rates. The gas inflow rates of galaxies decline by $\approx0.6 - 0.4$ dex in $\sim1.5$ Gyrs once they become bound to the massive cluster halo. In contrast, after infall into their respective host halo at $z=2$, field satellites continue to have comparable gas inflow rates with the field centrals (yellow dashed line; Figure \ref{fig:infall_inflow} right panel). The lack of decline in gas inflow rates in the field satellites suggests that the environment affects the gas inflow rate of galaxies only in massive cluster halos (median $\log \rm{M_{halo}/M_\odot} = 13.3 \pm 0.03$).

\subsection{Gas Inflow Rates vs Host Halo Mass }
\label{sec:halom}
In Figure \ref{fig:gas_in}, we show that the cluster centrals ($\rm{M}_{halo} \geq 10^{13} \rm{M}_\odot$) have higher ($\approx 0.5$ dex) gas inflow rates compared to the field centrals ($\rm{M}_{halo} < 10^{13} \rm{M}_\odot$) at $z=2$. In contrast, the satellite sample show an inverse relation at $z=2$ with the field satellites having higher gas inflow rates than the cluster satellites. In this section, we explore the correlation between gas inflow rates of galaxies with their host halo mass at different redshifts.

\begin{figure*}
    \begin{center}
    \includegraphics[width = \textwidth]{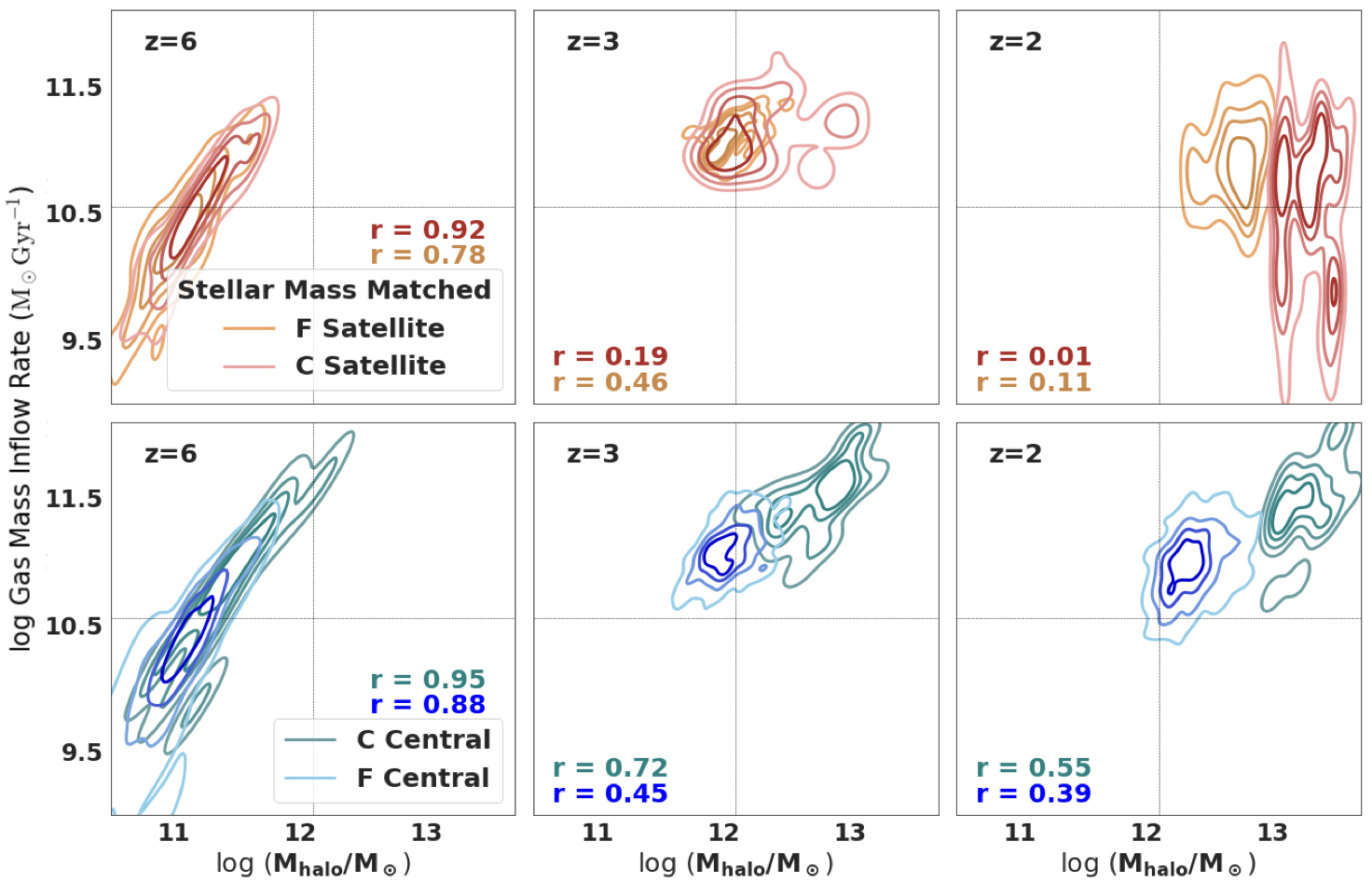}
    \end{center}
    
     \caption{ History of gas inflow rates of field and cluster satellite galaxies (top panel: shown in yellow and red contours respectively) and central galaxies (lower panel: shown in blue and teal contours respectively) elected at $z=2$ vs parent halo mass of the galaxy at redshifts  6, 3 and  2 (left to right). The gas inflow rates of both cluster and field satellites and centrals show a significant positive correlation with the halo mass of the host halo at $z=6$. However by $z<=3$, gas inflow rates of satellite galaxies have no significant dependence on the host halo mass of the galaxy, whereas centrals continue to have a positive correlation with the halo mass. The crosshairs help visually compare the gas inflow rates - host halo mass plane relation for the different populations at each redshift snapshot. }
     
    \label{fig:gas_halo_cen}
\end{figure*}

Figure \ref{fig:gas_halo_cen} shows the history  of gas inflow rates of field centrals and cluster centrals selected at $z=2$. The gas inflow rates are presented at $z=6,3$ and 2 against the host halo mass of the galaxies at respective redshifts. We calculate the Spearman correlation coefficient (r) to quantify any correlation between the gas inflow rates of the galaxies and their host halo masses (Table \ref{table:corr}). Gas inflow rates in cluster and field centrals show a very strong correlation with the host halo mass in the early universe until $z=3$, after which the strength of the correlation is moderate by $z=2$ (Table \ref{table:corr}).

Throughout the history ($6>z>2$) of central galaxies, the cluster centrals occupy higher halo masses (median $\log \rm{M_{halo}/M_\odot} = 13.2 \pm 0.03$ at $z=2$) and have higher correlation between gas inflow rates and their host halo masses compared to the field centrals (median $\log \rm{M_{halo}/M_\odot} = 12.3 \pm 0.02$ at $z=2$). The high value of r indicates that the gas inflow rates of centrals are dependent on their host halo mass at $6>z>2$, and the strength of the correlation at each redshift increases with increasing halo mass (Table \ref{table:corr}).

Figure \ref{fig:gas_halo_cen} shows the history of gas inflow rates of stellar mass matched field satellites and cluster satellites at redshifts 6, 3 and 2. In the early universe at $z=6$, we find a strong positive correlation between the gas inflow rates in galaxies and their respective host halo masses in both cluster satellites and field satellites (Table \ref{table:corr}). At $z=6$, the cluster and field satellites are associated with halos of similar halo masses and have comparable gas inflow rates. By $z=3$, the gas inflow rates in field satellites and their host halo mass show moderate correlation, in contrast, the cluster satellites show no significant correlation (Table \ref{table:corr}). By $z = 2$, both field and cluster satellites show no significant correlation (Table \ref{table:corr}) between the gas inflow rates and their host halo mass.

We show in Figure \ref{fig:gas_halo_cen} and Section \ref{sec:bi_ai}, that a population of cluster satellites have lower gas inflow rates compared to the field centrals at $z=2$. The lack of correlation between the gas inflow rate in the cluster satellites with their host halo mass indicate that the lower gas inflow rates in this population of cluster satellites is not driven by the host halo mass. We investigate the cause of low gas inflow rate in this population in Section \ref{sec:halor}.

\begin{table*}

    \begin{threeparttable}

        \caption{Correlation Coefficient between Gas inflow rates and Galaxy Properties}
        \label{table:corr}
        \begin{tabular}{lllllllll}
        \hline
        Sample &   \shortstack{ $\rm{M_*}$ \\ \small{(z=2)}}  & \shortstack{ $\rm{M_*}$ \\ \small{(z=3)}}&  \shortstack{$\rm{M_*}$ \\ \small{(z=6)}} & \shortstack{$\rm{M_{halo}}$\\ \small{(z=2)}} & \shortstack{$\rm{M_{halo}}$\\ \small{(z=3)}} & \shortstack{$\rm{M_{halo}}$ \\ \small{(z=6)}} & \shortstack{$\rm{R/R}_{200}$ \\ \small{(z=2)}}& \shortstack{$\log D_5$ \tnote{a}\\ \small{(z=2)}} \\   
        \hline \hline
        F Satellites  & 0.10 & 0.36 & 0.10 & 0.11 & 0.46 & 0.78 & 0.36 & 0.0 \\
        F Centrals    & 0.11 & 0.13 & 0.11 & 0.39 & 0.45 & 0.88 & - & $-0.40$\\
        C Satellites  & 0.16 & $-0.17$ & 0.16 & 0.01 & 0.19 & 0.92 & 0.54 & 0.27\\
        C Centrals    & - & - & - & 0.55 & 0.72 & 0.95 & - & $-0.67$ \\
        \hline
        \end{tabular}

        \begin{tablenotes}
        \item[a] $\log D_5$ is inversely proportional to local galaxy overdensity, hence, a negative value of correlation coefficient (r) implies a positive correlation of gas inflow rate with the local galaxy overdensity. 
        \end{tablenotes}
    \end{threeparttable}

\end{table*}

\subsection{Gas Inflow Rates vs Halo-centric Radius}
\label{sec:halor}
We show in section \ref{sec:bi_ai} and Figure \ref{fig:infall_inflow} that after infall, the gas inflow rates decrease rapidly in satellites of cluster halos only. However, the gas inflow rates for satellite galaxies at $z=2$ do not depend on the host halo mass. To resolve the seemingly contradictory results, we consider the correlation of gas inflow rates and the halo-centric distance. 

\begin{figure*}
    \begin{center}
    \includegraphics[width = \textwidth]{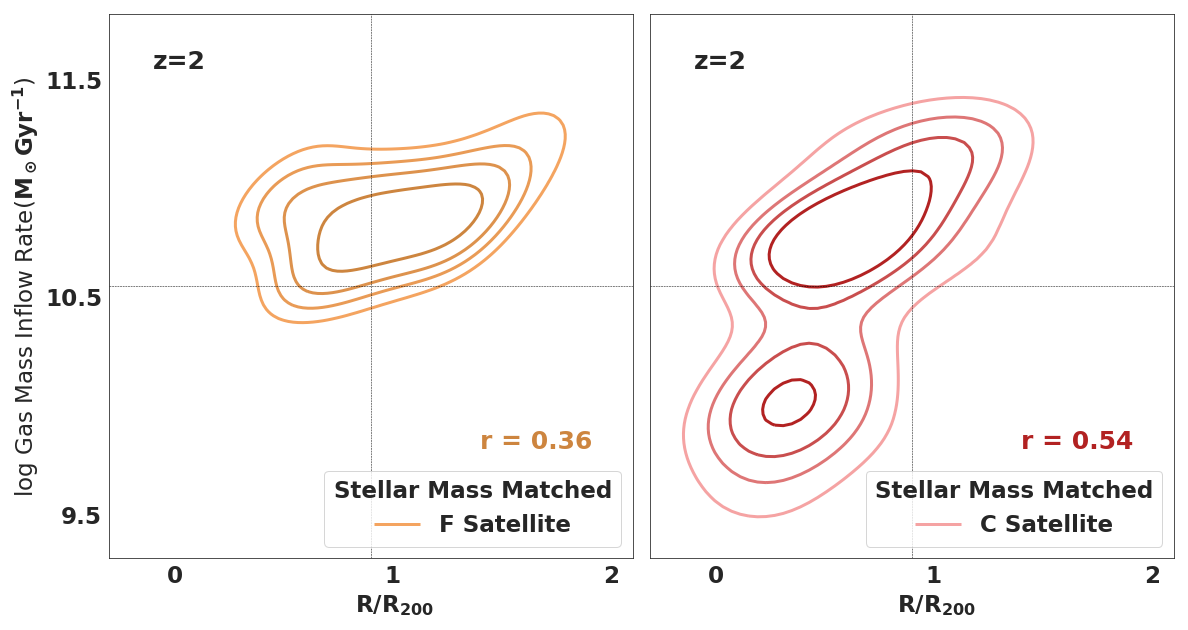}
    \end{center}
    
    \caption{ The gas inflow rate of cluster satellites (red contours; right panel) show a positive moderate correlation with the halo-centric radii normalized by $\rm{R}_{200}$ of the host halo at $z=2$. In contrast, the gas inflow rate of field satellites (yellow contours; left panel) shows a weak correlation with the halo-centric distance. The contours show the distribution of stellar mass matched sample of field satellite and cluster satellite galaxies on gas mass inflow rates - halo-centric radii plane respectively. The Correlation coefficient (r) for each sample is shown in respective panels. }
    \label{fig:dist}
\end{figure*}

Figure \ref{fig:dist} shows gas inflow rates of the stellar mass matched field satellites and the cluster satellites as a function of their halo-centric distance at $z=2$. We calculate the proper distance (R) of galaxies from the halo center (defined by the position of the particle with minimum gravitational potential energy) and normalize it by $\rm{R}_{200}$ (radius of a sphere centered at the halo center where mean density is 200 times the critical density of the Universe) at $z=2$. We calculate the Spearman correlation coefficient (r) between the gas inflow rates of satellite galaxies and their halo centric distance  $\rm{R/R}_{200}$ and report in Table (\ref{table:corr}). 

The gas inflow rate is weakly correlated to $\rm{R/}R_{200}$ for field satellites and moderately correlated for cluster satellites (Table \ref{table:corr}). The stellar mass matched cluster satellites occupy the halo closer to the cluster center (median $\rm{R/R}_{200} = 0.5\pm 0.2$) and have lower gas inflow rates compared to the field satellites (median $\rm{R/R}_{200} = 0.9 \pm 0.1$). This indicates that after galaxies fall into the massive cluster halos, their gas inflow rates decrease as they approach the halo center (Table \ref{table:inflow}).

\subsection{Gas Inflow Rates vs Local Galaxy Overdensity}
\label{sec:overd}
\begin{figure*}
    \begin{center}
    \includegraphics[width = \textwidth]{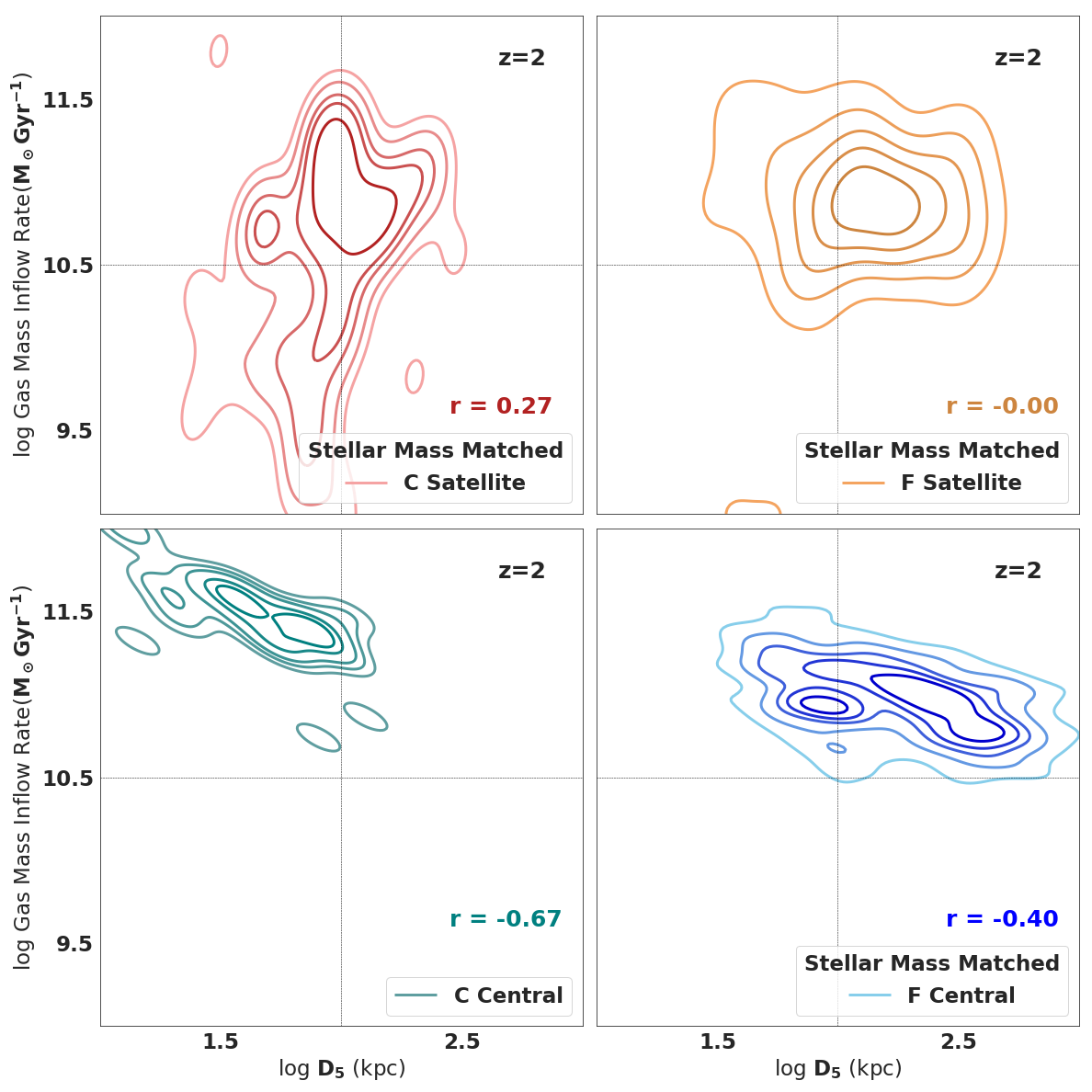}
    \end{center}
    
    \caption{ Gas inflow rate vs local galaxy overdensity (quantified as 5th nearest neighbor distance; $\rm{D_5}$) for stellar mass matched cluster satellites (red contours; left top), field satellites (yellow contours; right top) and field centrals (blue contours; right bottom) at $z=2$. The teal contours show the gas inflow rates of cluster centrals (not stellar mass matched) at $z=2$. The correlation coefficient (r) for each sample is shown in respective panels. The gas inflow rates in cluster and field centrals show a positive correlation with the local galaxy overdensity. The cluster satellites show a weak negative correlation with the local galaxy overdensity. In contrast, field satellites show no correlation with the local galaxy overdensity.}
    \label{fig:local_over}
\end{figure*}

In this section, we investigate the effect of local galaxy overdensity on the gas inflow rates of galaxies. We calculate the local galaxy overdensity by measuring the proper distance of the fifth nearest neighbor ($\rm{D_5}$; neighbor \logMstarMsun $\geq8$) for each galaxy at $z=2$. Figure \ref{fig:local_over} shows gas inflow rates of the four sample sets as a function of the fifth nearest neighbor distance.

We calculate the Spearman correlation coefficient (r) between the gas inflow rates and the fifth nearest neighbor distance ($\rm{D_5}$). The gas inflow rates in cluster and field centrals are positively correlated with the local galaxy overdensity at $z=2$ (Table \ref{table:corr}). The cluster centrals are in higher density environments (median $\log (\rm{D}_5 /\rm{kpc}) = 1.6 \pm 0.1 $ )  and have higher gas inflow rates compared to field centrals (median $\log (\rm{D}_5 /\rm{kpc}) = 2.2 \pm 0.02 $ ; Table \ref{table:inflow}). This indicates an effect of local galaxy overdensity on the gas inflow rates of centrals.

The cluster satellite galaxies show a weak negative correlation between gas inflow rates and the local galaxy overdensity. However, gas inflow rates in field satellites show no correlation with the local galaxy overdensity. The cluster satellite galaxies are in higher density environments (median $\log (\rm{D}_5 /\rm{kpc}) = 1.9 \pm 0.04 $) and have lower gas inflow rates compared to field satellites (median $\log (\rm{D}_5 /\rm{kpc}) = 2.0 \pm 0.1 $). Contrary to the effect on centrals, the local galaxy overdensity has a negative effect on the gas inflow rates of satellite galaxies.

\section{Discussion}
\label{sec:discussion}

Gas inflow in galaxies is expected to depend on the galaxy environment. We use \illustris\ cosmological simulations to study the effect of environment on the gas inflow rates in massive galaxies (\logMstarMsun $\geq 10.5$) at $z\geq2$. We select central and satellites galaxy samples in low mass (field; $\rm{M}_{halo} < 10^{13} \rm{M}_\odot$) and high mass (cluster; $\rm{M}_{halo} \geq 10^{13} \rm{M}_\odot$) halos at $z=2$. Furthermore, we have stellar mass matched the field satellite and centrals with the cluster satellite sample. The same can not be done for the cluster centrals because they are $\approx 4$ times more massive (median \logMstarMsun $= 11.2\pm0.2$) compared to the cluster satellite galaxies (median \logMstarMsun $= 10.6 \pm 0.1$). Our analysis supports a picture where gas inflow rates vary as a function of environment as characterized by mass of the host halo, halo-centric radius, and local galaxy overdensity.

\subsection{Cluster vs Field}

Satellite galaxies in cluster halos selected at $z=2$ have high gas inflow rates compared to the field centrals before infall into the host halo at $z=2$ (solid red line in Figure \ref{fig:infall_inflow}). After the infall into massive cluster halos, gas inflow rates in cluster satellites rapidly decline (dashed red line in Figure \ref{fig:infall_inflow}). However, the same trend is not observed in the field satellite galaxies selected at $z=2$, that show comparable gas inflow rates with the field centrals before and after infall (yellow solid and dashed lines in Figure \ref{fig:infall_inflow}). 

Similar to our result with \illustris, studies using other simulations like \cite{Simha2009} find that gas accretion rates in satellite galaxies at $z=0$ have declined to zero within 0.5-1 Gyrs after infall. At $z=2$, we find that decline in gas inflow rates in cluster satellite galaxies after infall is $\approx 0.5$ dex in 1 Gyr. In agreement with our results, \cite{Simha2009} and \cite{VandeVoort2017a} also find a divergence in gas inflow rates of satellite sample from the stellar mass matched "control" central sample only when the satellite sample are residing in more massive halos.

Before infall into the host halo at $z=2$, the massive satellite galaxies could be the centrals of their host halos. At $z=6$-4, cluster satellites have high gas inflow rates (Figures \ref{fig:gas_in} and \ref{fig:infall_inflow}) and are in more massive host halos compared to the field satellites (Figure \ref{fig:gas_halo_cen}). As the gas inflow rates of centrals at redshifts 6 to 2 is driven by their halo mass, we conclude that the higher gas inflow rates in the cluster satellites at redshifts 6 to 4 is also driven by their higher halo mass.

In section \ref{sec:halor}, we show the dependence of gas inflow rates of cluster and field satellite galaxies on the halo centric distance. Both cluster and field satellites (stellar mass matched) have lower gas inflow rates as they approach the halo center. The trend is stronger in cluster satellites compared to the field satellites as they are also in more overdense environment (Figure \ref{fig:local_over}). We also find that the massive field satellites are in the outskirts (median $ \rm{R}/{R}_{200} = 0.9 \pm 0.1$) of the halo, whereas the massive cluster satellites are closer (median $ \rm{R/R}_{200} = 0.5 \pm 0.2$) to the halo center (Figure \ref{fig:dist}; Table \ref{table:sample}). The cluster satellites are nearly a factor of two closer to their central galaxy compared to the field satellites. Thus it is likely that the cluster satellites have been part of their parent halo for longer compared to the field satellites. This could drive the lower gas inflow rates in cluster satellites compared to the field satellites.


\subsection{Centrals vs Satellites}

The gas inflow rates in satellite galaxies decreases with decreasing halo-centric distance (Figure \ref{fig:dist}). Cluster satellites within $\sim R_{200}$ radii have lower gas inflow rates and drive the decline in the median gas inflow rates of the entire cluster satellite sample at $z=2$ (median $ \rm{R/R}_{200} = 0.5 \pm 0.2$, Figure \ref{fig:infall_inflow}). Similar decline in gas inflow rates is not measured in the field satellite sample because they are in the outskirts of the halo (median $ \rm{R}/{R}_{200} = 0.9 \pm 0.1$; Figure \ref{fig:dist}). Gas inflow rates in both cluster and field satellites show a significant correlation with their halo and stellar mass at $z>3$ (Figures \ref{fig:gas_halo_cen} and \ref{fig:gas_sm_bi}). At $z>3$, our satellite sample has not fallen into the cluster potentials they are associated to at $z=2$. Thus before infall, these massive galaxies, could be the centrals of their parent halos at $z>3$ and their gas inflow rates are driven by their host halo mass.

Contrary to the satellites, gas inflow rates in central galaxies of both cluster and field halos have a strong dependence on the host halo mass (Figure \ref{fig:gas_halo_cen}) and local galaxy overdensity (Figure \ref{fig:local_over}). \cite{VandeVoort2017} and \cite{Simha2009} also find comparable results. The opposite correlation of the gas inflow rate with galaxy overdensity between cluster centrals and satellites could also indicate two different modes of gas accretion. While centrals are being fed by the filaments in the cluster center increasing gas accretion, the gas around the satellite galaxies in higher overdensities are unable to cool efficiently and hence hinder gas accretion \citep{voort2012,Chartab2021}.

Gas inflow rates of the field centrals are correlated with their stellar mass at redshifts 6 to 3, but not after $z\leq3$ (appendix \ref{appendix}). However, we find that the gas inflow rates in the field centrals are correlated with their halo mass for longer (at redshifts 6 to 2) (Figure \ref{fig:gas_halo_cen}). Thus we infer that the gas inflow rates in the centrals is driven mainly by their host halo mass and not their stellar mass.

The emerging scenario from our results is that after infall, satellites galaxies in massive cluster halos compete for gas with the cluster halo and the central galaxy (closer to the lowest point in halo potential and higher local galaxy overdensity). Therefore, cluster satellites are not able to accrete gas as they approach the halo center in massive halos. The gas inflow rates in satellite galaxies in field halos (that tend to be in the outskirts of the halo) are not affected by the environment (Figure \ref{fig:local_over}; Table \ref{table:sample}).

\subsection{Comparing Gas Fractions and Gas Inflow Rates from Observations to \illustris}

In \cite{Harshan2021a}, we find that the total gas fraction is lower only in the most massive protocluster galaxies compared to the field galaxies. In this work, we show that the total gas inflow rates (primordial +mergers) in the cluster satellite galaxies is lower than the field counterparts at $z=2$. We also find that the cluster satellites have a higher gas inflow rates compared to their field counterparts until $z\gtrsim 4$ before their infall into the cluster potential (driven by their halo mass, Figure \ref{fig:gas_halo_cen}). This indicates that in the formative stages of the clusters, satellite galaxies have higher gas inflow rates than field galaxies. This is consistent with observational studies that find a higher molecular gas fraction in galaxies in the protocluster environment \citep{Cucciati2014, Noble2017a, Rudnick2017a} at $z\sim2$. 

Direct detection of gas inflow rates continues to be elusive in observations at high redshifts and relies heavily on $\rm{Ly}\alpha$ emission, CGM absorption studies with quasar sightlines and mass-metallicity relation. In the past few decades an increasing number of proto-cluster/ filament galaxies at $z>2$ with a higher frequency of $\rm{Ly}\alpha$ emission line detections have been made \citep{Cantalupo2014, Hennawi2015, Umehata2019, Daddi2021}. Although $\rm{Ly}\alpha$ emission is used to identify gravitationally driven gas inflows, the interpretation of $\rm{Ly}\alpha$ emission is not without degeneracies. Often, $\rm{Ly}\alpha$ emission can be explained by outflows and photo-ionization by star formation and quasar. Within the interpretation of $\rm{Ly}\alpha$, studies find that at $2<z<3.3$ galaxies in filaments, galaxy groups or protoclusters \citep[$12.8<\log \rm{M_{halo}/M_\odot}<14.2$]{Daddi2021},\citep[$\log \rm{M_{halo}/M_\odot}\sim 15.4$]{Martin2014,Umehata2019}  have active cold gas inflows. In context of our results from \illustris, higher gas inflow rates in protocluster and groups indicate early stage of cluster formation.

Previous observational studies also investigate the effect of environment on gas accretion using the mass-metallicity relations and find conflicting results. At $z\sim2$ studies like \cite{Kacprzak2015,Alcorn2019a} find no significant difference between the metallicities of protocluster ($\log \rm{M_{halo}/M_\odot}\sim14.4$) and field galaxies, \cite{Shimakawa2015c}  find higher metallicities in protocluster ($\log \rm{M_{halo}/M_\odot}\sim14$) and hypothesise that the enhancement of metallicity in protocluster galaxies is due to the gas stripping. \cite{Chartab2021} find relative metallicity of galaxies in overdense regions reverses between redshifts 1.5 and 2. At $z\sim2$ galaxies in high density ($1+\delta > 1.77$) environment are relatively metal poor, potentially due to the effective cooling and/or higher gas inflow rates in the early phases of cluster assembly. However, by $z\sim1.5$ galaxies in high density environment ($1+\delta > 1.69$) becomes metal-rich. Our results suggest that as galaxies fall into the higher density environment, their gas inflow rates decline hence increasing their metallicities. Hence the contradicting observational results could be explained by different stages of protocluster assembly or selection of central and satellite galaxies. We also note that metallicity indicators using nebular emission lines trace the ionized ISM and hence may not be robust means of interpreting primordial gas inflows.

\subsection{Gas Inflow and Galaxy Quenching}

In \cite{Harshan2021a}, by constructing star formation histories using SED fitting of multi-wavelength data (0.1 - 1 microns), we find that the massive galaxies in the proto-cluster environment have formed $45 \pm8\%$ of their total stellar mass by $z=3$ compared to field galaxies that have formed $31\pm2 \%$ of their total stellar mass. Using \illustris, we find that the difference in star formation histories could be driven by the lower total gas fraction in the progenitors of massive cluster galaxies compared to the field counterparts. This indicates that the selected massive protocluster galaxies had lower gas inflows (from primordial inflow + mergers) at $z\lesssim5 $. In context of the results presented in this paper, we can confirm that the lower gas fraction of proto-cluster galaxies is in part due to lower gas inflow rates in cluster satellite galaxies. Thus our results suggest that the  observed onset of decline in star formation rates in $z=2$ massive protoclusters galaxies is at least in part driven by gas starvation. This early onset of environmental quenching in proto-cluster galaxies at $z=2$ could contribute to the large population of massive quenched galaxies in clusters at $z<0.5$.

The gas inflow rate of the cluster satellites increases at redshifts 6 to 4 but starts declining after their infall into the cluster halo at $z~4$. On the other hand, the gas inflow rate in the cluster central increases until $z~2$. This indicates that as massive cluster satellites are falling into the cluster potential, the gas that was available for accretion at redshifts 6 to 4 prior to infall, is now being redirected to the cluster potential and the cluster centrals, increasing their gas inflow rates. This diversion of gas could be due to the higher gravitational potential of the cluster centrals and the cluster halo. Thus cluster centrals have a higher constant gas inflow rates at redshifts 3 to 2, compared to the gas inflow rates in cluster satellites. The gas that is accreted onto the cluster halo would eventually build the hot ICM that is observed in $z>1$ clusters \citep{Rabitz2017, Stanford2001} and further impact the gas accretion in satellite galaxies in cluster halos. The decreased gas inflow rates after infall in addition to the interaction with the ICM will further decrease the star formation in the satellite galaxies building up the quenched cluster population observed at $z<0.5$. We will extend our work and investigate the cause of constant gas inflow rates in centrals (a redshifts 4 - 2) and the effect of the ICM on gas inflow rates and star formation in a future study.

\section{Summary}
\label{sec:summary}

In \cite{Harshan2021a}, we find that the lower star formation in massive protocluster galaxies compared to the field galaxies at $z=2$ is driven by lower total gas fractions since $z\sim5$. To investigate the cause of lower total gas fractions in protocluster galaxies, we study gas inflow rates. In this work, we use \illustris\ cosmological simulations to understand the role of environment in driving the gas inflow rates in the massive (\logMstarMsun $>10.5$) galaxies. We divide our sample into centrals and satellites in cluster ($\rm{M}_{halo} \geq 10^{13} \rm{M}_\odot$) and field ($\rm{M}_{halo} < 10^{13} \rm{M}_\odot$) halos. Our main results are summarised as follows: 

1) History of gas inflow rates at $6>z>2$: Until $z=2$, gas inflow rates in cluster centrals is higher compared to the field centrals. At $z=2$, gas inflow rates in cluster centrals (\logMgasin\, $=11.4\pm 0.3$ \inflow) are $\approx 0.5$ dex higher compared to field centrals which are not stellar mass matched (\logMgasin \,$= 10.9 \pm 0.01$ \inflow). The cluster satellites also have higher gas inflow rates compared to the stellar mass matched field centrals at $6>z>\sim4$ (Figure \ref{fig:gas_in}). At $z<\sim3$, the gas inflow rates in cluster satellites rapidly declines (by $\approx$ 0.4-0.6 dex) as they fall into the cluster potential. However, the history of gas inflow rates in field satellite galaxies do not show this trend and closely follow that of the field centrals (Figure \ref{fig:infall_inflow}). 


2) Gas inflow rates as a function of halo mass: At $z>4$, we find a strong correlation between the gas inflow rates of both cluster and field satellites, which is not observed at $z<4$. By $z=2$, the gas inflow rates of the cluster satellites decline compared to the field satellites (Figure \ref{fig:gas_halo_cen}). On the other hand, gas inflow rates of the central galaxies are strongly correlated with the host halo mass until $z=2$ (Figure \ref{fig:gas_halo_cen}). This indicates that satellites are competing for gas accretion with the centrals that have an advantageous position in the halo. 

3) Gas inflow rates as a function of halo-centric distance: Gas inflow rates of cluster satellites show a moderate correlation with the halo-centric radius such that the inflow rates decrease as galaxies approach the cluster center. A weak correlation between gas inflow rate and halo-centric distance is seen in the field satellites. Massive field satellites are also on average in the outskirts of the field halo, unlike the cluster satellites. This indicates that the field satellites have not been associated with the field halo for long unlike the cluster satellites and could explain the lack of decline in gas inflow rates in field satellites.

4) Gas inflow rates as a function of local galaxy overdensity: Gas inflow rates of central galaxies in both field and cluster halos are positively correlated with the local galaxy overdensity. Cluster centrals that live in higher mass halos and higher local galaxy overdensity have significantly higher gas inflow rates compared to the field centrals. Thus for centrals, gas inflow rates are primarily governed by the host halo mass and the local galaxy overdensity at $z=2$. However, the gas inflow rates in cluster satellites only show a weak negative correlation with the local galaxy overdensity and the field satellites show no correlation with the local galaxy overdensity. This indicates that for satellite galaxies the local galaxy overdensity have minimal or no effect on the gas inflow rates. 

We show that the effect of environment on the evolution of galaxies becomes significant in the very early universe (at $z\sim$ 3-4). Unlike the effects of environment in the local universe which are driven by galaxy-galaxy interactions or hydrodynamical interactions with the intracluster medium, in the early universe, the effect of environment on galaxies is driven by the position of the galaxies in the cosmic web (e.g., central vs satellite and high vs low host halo mass). While the gas inflow rates in centrals are driven mainly by the host halo mass and the local galaxy overdensity, gas inflow rates for satellite galaxies is driven by the halo-centric distance and the time since infall. 

The rapid decline in gas inflow in cluster satellites and the high gas inflow in cluster centrals indicate that the gas is redirected from the satellites to the cluster halo. This redirected gas should build up (starting as early as $z\sim3$) the ICM and is also accreted onto the cluster centrals. Whereas the cluster satellites with declining gas inflow rates become gas starved. Thus the quenching of the massive cluster satellites driven by starvation as shown in this work would lead to the build up of the massive quenched cluster populations that is observed in the local universe.

Further investigation of the inflowing gas by distinguishing the multi-phase gas and studying the evolution to $z=0$ is required to understand the build up of ICM and quenching in cluster galaxies driven by gas starvation. While the observational signature for gas accretion remains elusive, future multi-wavelength observational campaigns producing deep spatially resolved spectroscopy of the high redshift galaxies will be able to test the predictions made in cosmological simulations.








\section*{Acknowledgements}

The authors would like to thank Dr. Annalisa Pillepich for insightful feedback for the paper. The authors acknowledge the support of the Australian Research Council Centre of Excellence for All Sky Astrophysics in 3 Dimensions (ASTRO 3D), through project
number CE170100013. 

\section*{Data Availability}

The TNG100 simulations are publicly available from the \illustris\ repository: https://www.tng-project.org. The processed data of this
article will be shared on reasonable request to the corresponding
author.



\bibliographystyle{mnras}

\begin{thebibliography}{}
\makeatletter
\relax
\def\mn@urlcharsother{\let\do\@makeother \do\$\do\&\do\#\do\^\do\_\do\%\do\~}
\def\mn@doi{\begingroup\mn@urlcharsother \@ifnextchar [ {\mn@doi@}
  {\mn@doi@[]}}
\def\mn@doi@[#1]#2{\def\@tempa{#1}\ifx\@tempa\@empty \href
  {http://dx.doi.org/#2} {doi:#2}\else \href {http://dx.doi.org/#2} {#1}\fi
  \endgroup}
\def\mn@eprint#1#2{\mn@eprint@#1:#2::\@nil}
\def\mn@eprint@arXiv#1{\href {http://arxiv.org/abs/#1} {{\tt arXiv:#1}}}
\def\mn@eprint@dblp#1{\href {http://dblp.uni-trier.de/rec/bibtex/#1.xml}
  {dblp:#1}}
\def\mn@eprint@#1:#2:#3:#4\@nil{\def\@tempa {#1}\def\@tempb {#2}\def\@tempc
  {#3}\ifx \@tempc \@empty \let \@tempc \@tempb \let \@tempb \@tempa \fi \ifx
  \@tempb \@empty \def\@tempb {arXiv}\fi \@ifundefined
  {mn@eprint@\@tempb}{\@tempb:\@tempc}{\expandafter \expandafter \csname
  mn@eprint@\@tempb\endcsname \expandafter{\@tempc}}}

\bibitem[\protect\citeauthoryear{{Alcorn} et~al.,}{{Alcorn}
  et~al.}{2019}]{Alcorn2019a}
{Alcorn} L.~Y.,  et~al., 2019, \mn@doi [\apj] {10.3847/1538-4357/ab3b0c}, \href
  {https://ui.adsabs.harvard.edu/abs/2019ApJ...883..153A} {883, 153}

\bibitem[\protect\citeauthoryear{{Angl{\'e}s-Alc{\'a}zar},
  {Faucher-Gigu{\`e}re}, {Kere{\v{s}}}, {Hopkins}, {Quataert}  \&
  {Murray}}{{Angl{\'e}s-Alc{\'a}zar} et~al.}{2017}]{Angles2017}
{Angl{\'e}s-Alc{\'a}zar} D.,  {Faucher-Gigu{\`e}re} C.-A.,  {Kere{\v{s}}} D.,
  {Hopkins} P.~F.,  {Quataert} E.,   {Murray} N.,  2017, \mn@doi [\mnras]
  {10.1093/mnras/stx1517}, \href
  {https://ui.adsabs.harvard.edu/abs/2017MNRAS.470.4698A} {470, 4698}

\bibitem[\protect\citeauthoryear{Aoyama, Kodama, Suzuki, Tadaki, Shimakawa,
  Hayashi, Koyama  \& P{\'{e}}rez-Mart{\'{i}}nez}{Aoyama
  et~al.}{2022}]{Aoyama2022}
Aoyama K.,  Kodama T.,  Suzuki T.~L.,  Tadaki K.-i.,  Shimakawa R.,  Hayashi
  M.,  Koyama Y.,   P{\'{e}}rez-Mart{\'{i}}nez J.~M.,  2022, \mn@doi [The
  Astrophysical Journal] {10.3847/1538-4357/ac34fa}, 924, 74

\bibitem[\protect\citeauthoryear{{Behroozi}, {Wechsler}, {Hearin}  \&
  {Conroy}}{{Behroozi} et~al.}{2019}]{Behroozi2019}
{Behroozi} P.,  {Wechsler} R.~H.,  {Hearin} A.~P.,   {Conroy} C.,  2019,
  \mn@doi [\mnras] {10.1093/mnras/stz1182}, \href
  {https://ui.adsabs.harvard.edu/abs/2019MNRAS.488.3143B} {488, 3143}

\bibitem[\protect\citeauthoryear{{Birnboim} \& {Dekel}}{{Birnboim} \&
  {Dekel}}{2003}]{Birnboim2003}
{Birnboim} Y.,  {Dekel} A.,  2003, \mn@doi [\mnras]
  {10.1046/j.1365-8711.2003.06955.x}, \href
  {https://ui.adsabs.harvard.edu/abs/2003MNRAS.345..349B} {345, 349}

\bibitem[\protect\citeauthoryear{{Bouch{\'e}}, {Murphy}, {Kacprzak},
  {P{\'e}roux}, {Contini}, {Martin}  \& {Dessauges-Zavadsky}}{{Bouch{\'e}}
  et~al.}{2013}]{Bouche2013}
{Bouch{\'e}} N.,  {Murphy} M.~T.,  {Kacprzak} G.~G.,  {P{\'e}roux} C.,
  {Contini} T.,  {Martin} C.~L.,   {Dessauges-Zavadsky} M.,  2013, \mn@doi
  [Science] {10.1126/science.1234209}, \href
  {https://ui.adsabs.harvard.edu/abs/2013Sci...341...50B} {341, 50}

\bibitem[\protect\citeauthoryear{Brown et~al.,}{Brown
  et~al.}{2017}]{Brown2017a}
Brown T.,  et~al., 2017, \mn@doi [Monthly Notices of the Royal Astronomical
  Society] {10.1093/mnras/stw2991}, 466, 1275

\bibitem[\protect\citeauthoryear{Cantalupo, Arrigoni-Battaia, Prochaska,
  Hennawi  \& Madau}{Cantalupo et~al.}{2014}]{Cantalupo2014}
Cantalupo S.,  Arrigoni-Battaia F.,  Prochaska J.~X.,  Hennawi J.~F.,   Madau
  P.,  2014, \mn@doi [Nature] {10.1038/nature12898}, 506, 63

\bibitem[\protect\citeauthoryear{{Chartab} et~al.,}{{Chartab}
  et~al.}{2021}]{Chartab2021}
{Chartab} N.,  et~al., 2021, \mn@doi [\apj] {10.3847/1538-4357/abd71f}, \href
  {https://ui.adsabs.harvard.edu/abs/2021ApJ...908..120C} {908, 120}

\bibitem[\protect\citeauthoryear{{Christopher Martin}, Chang, Matuszewski,
  Morrissey, Rahman, Moore, Steidel  \& Matsuda}{{Christopher Martin}
  et~al.}{2014}]{Martin2014}
{Christopher Martin} D.,  Chang D.,  Matuszewski M.,  Morrissey P.,  Rahman S.,
   Moore A.,  Steidel C.~C.,   Matsuda Y.,  2014, \mn@doi [The Astrophysical
  Journal] {10.1088/0004-637X/786/2/107}, 786, 107

\bibitem[\protect\citeauthoryear{Chun, Smith, Shin, Kim  \& Raouf}{Chun
  et~al.}{2020}]{Chun2020}
Chun K.,  Smith R.,  Shin J.,  Kim S.~S.,   Raouf M.,  2020, \mn@doi [The
  Astrophysical Journal] {10.3847/1538-4357/ab5afb}, 889, 173

\bibitem[\protect\citeauthoryear{Cortese, Catinella  \& Smith}{Cortese
  et~al.}{2021}]{Cortese2021}
Cortese L.,  Catinella B.,   Smith R.,  2021, \mn@doi [Publications of the
  Astronomical Society of Australia] {10.1017/pasa.2021.18}, 38

\bibitem[\protect\citeauthoryear{Cucciati et~al.,}{Cucciati
  et~al.}{2014}]{Cucciati2014}
Cucciati O.,  et~al., 2014, \mn@doi [Astronomy & Astrophysics]
  {10.1051/0004-6361/201423811}, 570, A16

\bibitem[\protect\citeauthoryear{Daddi et~al.,}{Daddi et~al.}{2021}]{Daddi2021}
Daddi E.,  et~al., 2021, \mn@doi [Astronomy and Astrophysics]
  {10.1051/0004-6361/202038700}, 649, 1

\bibitem[\protect\citeauthoryear{Daddi et~al.,}{Daddi et~al.}{2022}]{Daddi2022}
Daddi E.,  et~al., 2022, \mn@doi [The Astrophysical Journal Letters]
  {10.3847/2041-8213/ac531f}, 926, L21

\bibitem[\protect\citeauthoryear{{Dijkstra} \& {Loeb}}{{Dijkstra} \&
  {Loeb}}{2009}]{Dijkstra2009}
{Dijkstra} M.,  {Loeb} A.,  2009, \mn@doi [\mnras]
  {10.1111/j.1365-2966.2009.15533.x}, \href
  {https://ui.adsabs.harvard.edu/abs/2009MNRAS.400.1109D} {400, 1109}

\bibitem[\protect\citeauthoryear{{Forbes} et~al.,}{{Forbes}
  et~al.}{2022}]{Forbes2022}
{Forbes} J.~C.,  et~al., 2022, arXiv e-prints, \href
  {https://ui.adsabs.harvard.edu/abs/2022arXiv220405344F} {p. arXiv:2204.05344}

\bibitem[\protect\citeauthoryear{Forrest et~al.,}{Forrest
  et~al.}{2017}]{Forrest2017a}
Forrest B.,  et~al., 2017, \mn@doi [The Astrophysical Journal]
  {10.3847/2041-8213/aa653b}, 838, L12

\bibitem[\protect\citeauthoryear{Fu, Xue, Prochaska, Stockton, Ponnada, Lau,
  Cooray  \& Narayanan}{Fu et~al.}{2021}]{Fu2021}
Fu H.,  Xue R.,  Prochaska J.~X.,  Stockton A.,  Ponnada S.,  Lau M.~W.,
  Cooray A.,   Narayanan D.,  2021, \mn@doi [The Astrophysical Journal]
  {10.3847/1538-4357/abdb32}, 908, 188

\bibitem[\protect\citeauthoryear{{Fumagalli}, {Krumholz}, {Prochaska},
  {Gavazzi}  \& {Boselli}}{{Fumagalli} et~al.}{2009}]{Fumagalli2009}
{Fumagalli} M.,  {Krumholz} M.~R.,  {Prochaska} J.~X.,  {Gavazzi} G.,
  {Boselli} A.,  2009, \mn@doi [\apj] {10.1088/0004-637X/697/2/1811}, \href
  {https://ui.adsabs.harvard.edu/abs/2009ApJ...697.1811F} {697, 1811}

\bibitem[\protect\citeauthoryear{Gunn \& Gott}{Gunn \& Gott}{1972}]{Gunn1972}
Gunn J.~E.,  Gott J. R.~I.,  1972, Astrophysical Journal, 176, 1

\bibitem[\protect\citeauthoryear{Gupta et~al.,}{Gupta et~al.}{2018}]{Gupta2018}
Gupta A.,  et~al., 2018, \mn@doi [Monthly Notices of the Royal Astronomical
  Society: Letters] {10.1093/mnrasl/sly037}, 477, L35

\bibitem[\protect\citeauthoryear{{Harshan} et~al.,}{{Harshan}
  et~al.}{2020}]{Harshan2020j}
{Harshan} A.,  et~al., 2020, \mn@doi [\apj] {10.3847/1538-4357/ab76cf}, \href
  {https://ui.adsabs.harvard.edu/abs/2020ApJ...892...77H} {892, 77}

\bibitem[\protect\citeauthoryear{{Harshan} et~al.,}{{Harshan}
  et~al.}{2021}]{Harshan2021a}
{Harshan} A.,  et~al., 2021, \mn@doi [\apj] {10.3847/1538-4357/ac0cf3}, \href
  {https://ui.adsabs.harvard.edu/abs/2021ApJ...919...57H} {919, 57}

\bibitem[\protect\citeauthoryear{Hayashi et~al.,}{Hayashi
  et~al.}{2017}]{Hayashi2017}
Hayashi M.,  et~al., 2017, \mn@doi [The Astrophysical Journal]
  {10.3847/2041-8213/aa71ad}, 841, L21

\bibitem[\protect\citeauthoryear{Hennawi, Prochaska, Cantalupo  \&
  Arrigoni-Battaia}{Hennawi et~al.}{2015}]{Hennawi2015}
Hennawi J.~F.,  Prochaska J.~X.,  Cantalupo S.,   Arrigoni-Battaia F.,  2015,
  \mn@doi [Science] {10.1126/science.aaa5397}, 348, 779

\bibitem[\protect\citeauthoryear{Kacprzak et~al.,}{Kacprzak
  et~al.}{2015}]{Kacprzak2015}
Kacprzak G.~G.,  et~al., 2015, \mn@doi [Astrophysical Journal Letters]
  {10.1088/2041-8205/802/2/L26}, 802

\bibitem[\protect\citeauthoryear{Kacprzak et~al.,}{Kacprzak
  et~al.}{2016}]{Kacprzak2016a}
Kacprzak G.~G.,  et~al., 2016, \mn@doi [The Astrophysical Journal]
  {10.3847/2041-8205/826/1/l11}, 826, L11

\bibitem[\protect\citeauthoryear{Kawinwanichakij et~al.,}{Kawinwanichakij
  et~al.}{2017}]{Kawinwanichakij2017b}
Kawinwanichakij L.,  et~al., 2017, \mn@doi [The Astrophysical Journal]
  {10.3847/1538-4357/aa8b75}, 847, 134

\bibitem[\protect\citeauthoryear{{Kere{\v{s}}}, {Katz}, {Weinberg}  \&
  {Dav{\'e}}}{{Kere{\v{s}}} et~al.}{2005}]{Keres2005}
{Kere{\v{s}}} D.,  {Katz} N.,  {Weinberg} D.~H.,   {Dav{\'e}} R.,  2005,
  \mn@doi [\mnras] {10.1111/j.1365-2966.2005.09451.x}, \href
  {https://ui.adsabs.harvard.edu/abs/2005MNRAS.363....2K} {363, 2}

\bibitem[\protect\citeauthoryear{{Kere{\v{s}}}, {Katz}, {Fardal}, {Dav{\'e}}
  \& {Weinberg}}{{Kere{\v{s}}} et~al.}{2009}]{Keres2009}
{Kere{\v{s}}} D.,  {Katz} N.,  {Fardal} M.,  {Dav{\'e}} R.,   {Weinberg} D.~H.,
   2009, \mn@doi [\mnras] {10.1111/j.1365-2966.2009.14541.x}, \href
  {https://ui.adsabs.harvard.edu/abs/2009MNRAS.395..160K} {395, 160}

\bibitem[\protect\citeauthoryear{{Liao} \& {Gao}}{{Liao} \&
  {Gao}}{2019}]{Liao2019}
{Liao} S.,  {Gao} L.,  2019, \mn@doi [\mnras] {10.1093/mnras/stz441}, \href
  {https://ui.adsabs.harvard.edu/abs/2019MNRAS.485..464L} {485, 464}

\bibitem[\protect\citeauthoryear{Marinacci et~al.,}{Marinacci
  et~al.}{2018}]{Marinacci2017}
Marinacci F.,  et~al., 2018, \mn@doi [Monthly Notices of the Royal Astronomical
  Society] {10.1093/mnras/sty2206}, 480, 5113

\bibitem[\protect\citeauthoryear{Moore, Katz, Lake, Dressler  \& Oemler}{Moore
  et~al.}{1996}]{Moore1996}
Moore B.,  Katz N.,  Lake G.,  Dressler A.,   Oemler A.,  1996, \mn@doi
  [Nature] {10.1038/379613a0}, 379, 613

\bibitem[\protect\citeauthoryear{{Moster}, {Naab}  \& {White}}{{Moster}
  et~al.}{2018}]{Moster2018}
{Moster} B.~P.,  {Naab} T.,   {White} S. D.~M.,  2018, \mn@doi [\mnras]
  {10.1093/mnras/sty655}, \href
  {https://ui.adsabs.harvard.edu/abs/2018MNRAS.477.1822M} {477, 1822}

\bibitem[\protect\citeauthoryear{Naiman et~al.,}{Naiman
  et~al.}{2018}]{Naiman2017}
Naiman J.~P.,  et~al., 2018, \mn@doi [Monthly Notices of the Royal Astronomical
  Society] {10.1093/mnras/sty618}, 477, 1206

\bibitem[\protect\citeauthoryear{Nanayakkara et~al.,}{Nanayakkara
  et~al.}{2016}]{Nanayakkara2016}
Nanayakkara T.,  et~al., 2016, \mn@doi [The Astrophysical Journal]
  {10.3847/0004-637X/828/1/21}, 828, 21

\bibitem[\protect\citeauthoryear{Nelson, Vogelsberger, Genel, Sijacki, Keres,
  Springel  \& Hernquist}{Nelson et~al.}{2013}]{Nelson2013}
Nelson D.,  Vogelsberger M.,  Genel S.,  Sijacki D.,  Keres D.,  Springel V.,
  Hernquist L.,  2013, \mn@doi [Monthly Notices of the Royal Astronomical
  Society] {10.1093/mnras/sts595}, 429, 3353

\bibitem[\protect\citeauthoryear{Nelson et~al.,}{Nelson
  et~al.}{2018}]{Nelson2018}
Nelson D.,  et~al., 2018, \mn@doi [Monthly Notices of the Royal Astronomical
  Society] {10.1093/mnras/stx3040}, 475, 624

\bibitem[\protect\citeauthoryear{Nelson et~al.,}{Nelson
  et~al.}{2019}]{Nelson2019a}
Nelson D.,  et~al., 2019, \mn@doi [Computational Astrophysics and Cosmology]
  {10.1186/s40668-019-0028-x}, 6, 2

\bibitem[\protect\citeauthoryear{Noble et~al.,}{Noble
  et~al.}{2017}]{Noble2017a}
Noble A.~G.,  et~al., 2017, \mn@doi [The Astrophysical Journal]
  {10.3847/2041-8213/aa77f3}, 842, L21

\bibitem[\protect\citeauthoryear{{Ocvirk}, {Pichon}  \& {Teyssier}}{{Ocvirk}
  et~al.}{2008}]{Ocvirk2008}
{Ocvirk} P.,  {Pichon} C.,   {Teyssier} R.,  2008, \mn@doi [\mnras]
  {10.1111/j.1365-2966.2008.13763.x}, \href
  {https://ui.adsabs.harvard.edu/abs/2008MNRAS.390.1326O} {390, 1326}

\bibitem[\protect\citeauthoryear{Peng, Maiolino  \& Cochrane}{Peng
  et~al.}{2015}]{Peng2015}
Peng Y.,  Maiolino R.,   Cochrane R.,  2015, \mn@doi [Nature]
  {10.1038/nature14439}, 521, 192

\bibitem[\protect\citeauthoryear{Pillepich et~al.,}{Pillepich
  et~al.}{2018a}]{Pillepich2018}
Pillepich A.,  et~al., 2018a, \mn@doi [Monthly Notices of the Royal
  Astronomical Society] {10.1093/mnras/stx2656}, 473, 4077

\bibitem[\protect\citeauthoryear{{Pillepich} et~al.,}{{Pillepich}
  et~al.}{2018b}]{Pillepich2018b}
{Pillepich} A.,  et~al., 2018b, \mn@doi [\mnras] {10.1093/mnras/stx3112}, \href
  {https://ui.adsabs.harvard.edu/abs/2018MNRAS.475..648P} {475, 648}

\bibitem[\protect\citeauthoryear{{Rabitz}, {Lamer}, {Schwope}  \&
  {Takey}}{{Rabitz} et~al.}{2017}]{Rabitz2017}
{Rabitz} A.,  {Lamer} G.,  {Schwope} A.,   {Takey} A.,  2017, \mn@doi [\aap]
  {10.1051/0004-6361/201731128}, \href
  {https://ui.adsabs.harvard.edu/abs/2017A&A...607A..56R} {607, A56}

\bibitem[\protect\citeauthoryear{{Rees} \& {Ostriker}}{{Rees} \&
  {Ostriker}}{1977}]{Rees1977}
{Rees} M.~J.,  {Ostriker} J.~P.,  1977, \mn@doi [\mnras]
  {10.1093/mnras/179.4.541}, \href
  {https://ui.adsabs.harvard.edu/abs/1977MNRAS.179..541R} {179, 541}

\bibitem[\protect\citeauthoryear{Rodriguez-Gomez et~al.,}{Rodriguez-Gomez
  et~al.}{2015}]{Rodriguez-Gomez2015}
Rodriguez-Gomez V.,  et~al., 2015, \mn@doi [Monthly Notices of the Royal
  Astronomical Society] {10.1093/mnras/stv264}, 449, 49

\bibitem[\protect\citeauthoryear{Rudnick et~al.,}{Rudnick
  et~al.}{2017}]{Rudnick2017a}
Rudnick G.,  et~al., 2017, \mn@doi [The Astrophysical Journal]
  {10.3847/1538-4357/aa87b2}, 849, 27

\bibitem[\protect\citeauthoryear{Scott, Usero, Brinks, Boselli, Cortese  \&
  Bravo-Alfaro}{Scott et~al.}{2012}]{Scott2013}
Scott T.~C.,  Usero A.,  Brinks E.,  Boselli A.,  Cortese L.,   Bravo-Alfaro
  H.,  2012, \mn@doi [Monthly Notices of the Royal Astronomical Society]
  {10.1093/mnras/sts328}, 429, 221

\bibitem[\protect\citeauthoryear{{Shimakawa}, {Kodama}, {Tadaki}, {Hayashi},
  {Koyama}  \& {Tanaka}}{{Shimakawa} et~al.}{2015}]{Shimakawa2015c}
{Shimakawa} R.,  {Kodama} T.,  {Tadaki} K.-i.,  {Hayashi} M.,  {Koyama} Y.,
  {Tanaka} I.,  2015, \mn@doi [\mnras] {10.1093/mnras/stv051}, \href
  {https://ui.adsabs.harvard.edu/abs/2015MNRAS.448..666S} {448, 666}

\bibitem[\protect\citeauthoryear{{Silk}}{{Silk}}{1977}]{Silk1977}
{Silk} J.,  1977, \mn@doi [\apj] {10.1086/154972}, \href
  {https://ui.adsabs.harvard.edu/abs/1977ApJ...211..638S} {211, 638}

\bibitem[\protect\citeauthoryear{Simha, Weinberg, Dav{\'{e}}, Gnedin, Katz  \&
  Kere}{Simha et~al.}{2009}]{Simha2009}
Simha V.,  Weinberg D.~H.,  Dav{\'{e}} R.,  Gnedin O.~Y.,  Katz N.,   Kere D.,
  2009, \mn@doi [Monthly Notices of the Royal Astronomical Society]
  {10.1111/j.1365-2966.2009.15341.x}, 399, 650

\bibitem[\protect\citeauthoryear{{Sparre}, {Whittingham}, {Damle}, {Hani},
  {Richter}, {Ellison}, {Pfrommer}  \& {Vogelsberger}}{{Sparre}
  et~al.}{2022}]{Martin2022}
{Sparre} M.,  {Whittingham} J.,  {Damle} M.,  {Hani} M.~H.,  {Richter} P.,
  {Ellison} S.~L.,  {Pfrommer} C.,   {Vogelsberger} M.,  2022, \mn@doi [\mnras]
  {10.1093/mnras/stab3171}, \href
  {https://ui.adsabs.harvard.edu/abs/2022MNRAS.509.2720S} {509, 2720}

\bibitem[\protect\citeauthoryear{Springel et~al.,}{Springel
  et~al.}{2018}]{Springel2018a}
Springel V.,  et~al., 2018, \mn@doi [Monthly Notices of the Royal Astronomical
  Society] {10.1093/mnras/stx3304}, 475, 676

\bibitem[\protect\citeauthoryear{{Stanford}, {Holden}, {Rosati}, {Tozzi},
  {Borgani}, {Eisenhardt}  \& {Spinrad}}{{Stanford}
  et~al.}{2001}]{Stanford2001}
{Stanford} S.~A.,  {Holden} B.,  {Rosati} P.,  {Tozzi} P.,  {Borgani} S.,
  {Eisenhardt} P.~R.,   {Spinrad} H.,  2001, \mn@doi [\apj] {10.1086/320583},
  \href {https://ui.adsabs.harvard.edu/abs/2001ApJ...552..504S} {552, 504}

\bibitem[\protect\citeauthoryear{Stern, Fielding, Faucher-Gigu{\`{e}}re  \&
  Quataert}{Stern et~al.}{2019}]{Stern2020}
Stern J.,  Fielding D.,  Faucher-Gigu{\`{e}}re C.-A.,   Quataert E.,  2019,
  \mn@doi [Monthly Notices of the Royal Astronomical Society]
  {10.1093/mnras/staa198}, 492, 6042

\bibitem[\protect\citeauthoryear{Straatman et~al.,}{Straatman
  et~al.}{2016}]{Straatman2016a}
Straatman C. M.~S.,  et~al., 2016, \mn@doi [The Astrophysical Journal]
  {10.3847/0004-637x/830/1/51}, 830, 51

\bibitem[\protect\citeauthoryear{Tran et~al.,}{Tran et~al.}{2010}]{Tran2010}
Tran K. V.~H.,  et~al., 2010, \mn@doi [Astrophysical Journal Letters]
  {10.1088/2041-8205/719/2/L126}, 719, 126

\bibitem[\protect\citeauthoryear{Tran et~al.,}{Tran et~al.}{2015}]{Tran2015}
Tran K.-V.~H.,  et~al., 2015, \mn@doi [The Astrophysical Journal]
  {10.1088/0004-637X/811/1/28}, 811, 28

\bibitem[\protect\citeauthoryear{Umehata et~al.,}{Umehata
  et~al.}{2019}]{Umehata2019}
Umehata H.,  et~al., 2019, \mn@doi [Science] {10.1126/science.aaw5949}, 366, 97

\bibitem[\protect\citeauthoryear{Watson et~al.,}{Watson
  et~al.}{2019}]{Watson2019}
Watson C.,  et~al., 2019, \mn@doi [The Astrophysical Journal]
  {10.3847/1538-4357/ab06ef}, 874, 63

\bibitem[\protect\citeauthoryear{Weinberger et~al.,}{Weinberger
  et~al.}{2017}]{Weinberger2017}
Weinberger R.,  et~al., 2017, \mn@doi [Monthly Notices of the Royal
  Astronomical Society] {10.1093/mnras/stw2944}, 465, 3291

\bibitem[\protect\citeauthoryear{Wright, Lagos, Power  \& Correa}{Wright
  et~al.}{2021}]{Wright2021}
Wright R.~J.,  Lagos C. D.~P.,  Power C.,   Correa C.~A.,  2021, \mn@doi
  [Monthly Notices of the Royal Astronomical Society] {10.1093/mnras/stab1057},
  504, 5702

\bibitem[\protect\citeauthoryear{Zabl et~al.,}{Zabl et~al.}{2019}]{Zabl2019}
Zabl J.,  et~al., 2019, \mn@doi [Monthly Notices of the Royal Astronomical
  Society] {10.1093/mnras/stz392}, 485, 1961

\bibitem[\protect\citeauthoryear{Zinger et~al.,}{Zinger
  et~al.}{2020}]{Zinger2020}
Zinger E.,  et~al., 2020, \mn@doi [Monthly Notices of the Royal Astronomical
  Society] {10.1093/mnras/staa2607}, 499, 768

\bibitem[\protect\citeauthoryear{van~de Voort \& Schaye}{van~de Voort \&
  Schaye}{2012}]{voort2012}
van~de Voort F.,  Schaye J.,  2012, \mn@doi [Monthly Notices of the Royal
  Astronomical Society] {10.1111/j.1365-2966.2012.20949.x}, 423, 2991

\bibitem[\protect\citeauthoryear{van~de Voort, Bah{\'{e}}, Bower, Correa,
  Crain, Schaye  \& Theuns}{van~de Voort et~al.}{2017a}]{VandeVoort2017}
van~de Voort F.,  Bah{\'{e}} Y.~M.,  Bower R.~G.,  Correa C.~A.,  Crain R.~A.,
  Schaye J.,   Theuns T.,  2017a, \mn@doi [Monthly Notices of the Royal
  Astronomical Society] {10.1093/mnras/stw3356}, 466, 3460

\bibitem[\protect\citeauthoryear{van~de Voort, Bah{\'{e}}, Bower, Correa,
  Crain, Schaye  \& Theuns}{van~de Voort et~al.}{2017b}]{VandeVoort2017a}
van~de Voort F.,  Bah{\'{e}} Y.~M.,  Bower R.~G.,  Correa C.~A.,  Crain R.~A.,
  Schaye J.,   Theuns T.,  2017b, \mn@doi [Monthly Notices of the Royal
  Astronomical Society] {10.1093/mnras/stw3356}, 466, 3460

\makeatother
\end{thebibliography}




\appendix

\section{Appendix - Gas Inflow Rates vs Stellar Mass of a Galaxy}
\label{appendix}
\begin{figure*}
    \begin{center}
    \includegraphics[width = \textwidth]{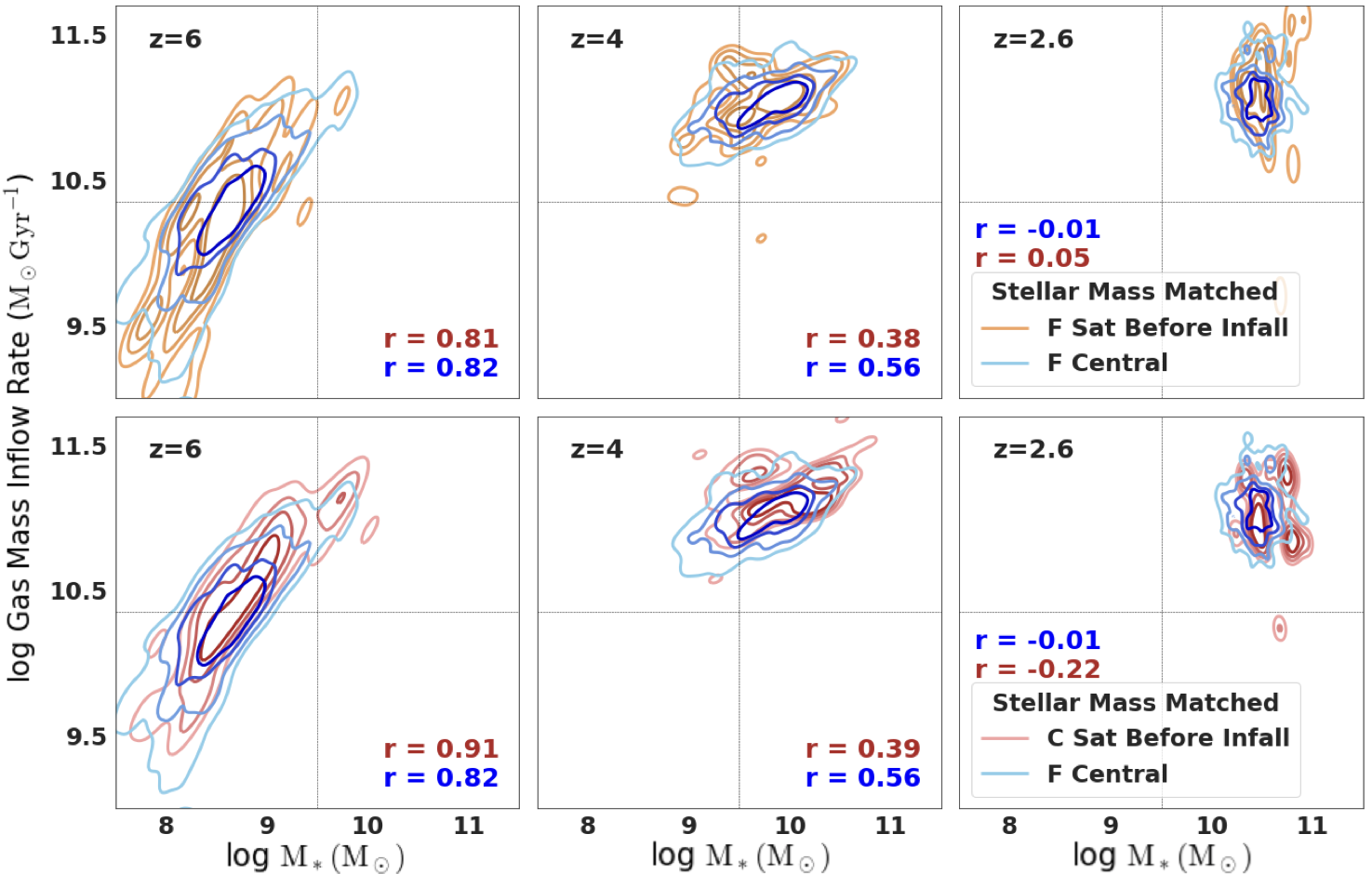}

    \end{center}
    
    \caption{Before infall, the gas inflow rates of stellar mass matched field (yellow contours; top row) and cluster satellite galaxies (red contours; bottom row) are comparable at redshifts  6,4 and 2.6 (left to right). The blue contours in both top and bottom rows show the comparison with the field centrals at the respective redshifts. Spearman correlation coefficient (r) between gas inflow rate and stellar mass of the galaxy is presented in the respective color in each panel. The crosshairs help visually compare the gas inflow rates for the different populations at each redshift snapshot. The positive correlation of gas inflow rates with the stellar mass of field and cluster satellites at $z>4$ vanishes at  $z\leq4$.}
    \label{fig:gas_sm_bi}
\end{figure*}

\begin{figure*}
    \begin{center}
    \includegraphics[width = \textwidth]{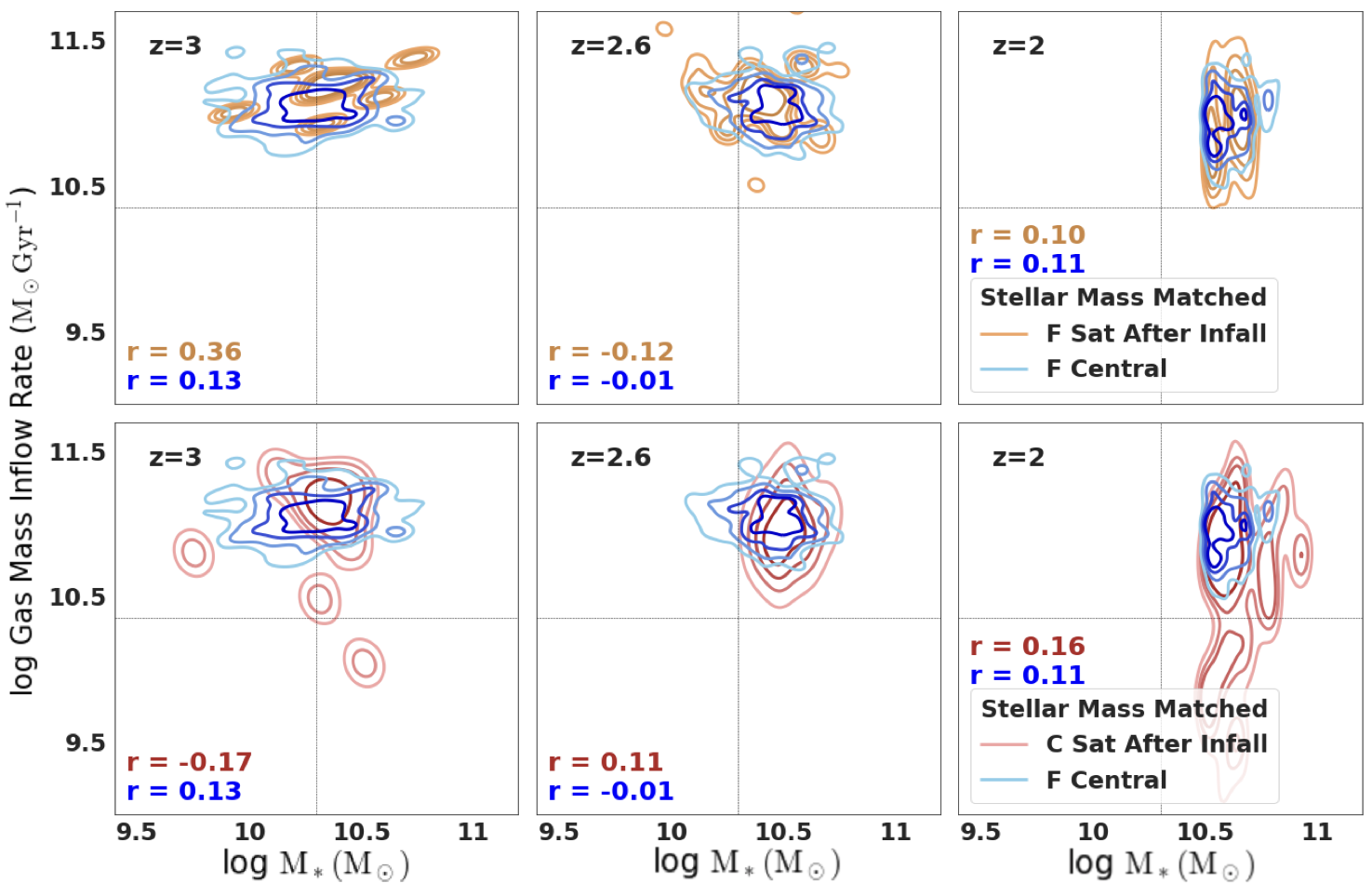}
    \end{center}
    
    \caption{After infall, the gas inflow rates of cluster satellites (red contours; bottom row) decline compared to the stellar mass matched field satellite sample (yellow contours; top row). The blue contours in both top and bottom rows show the comparison with the field centrals at the respective redshifts. Spearman correlation coefficient (r) between gas inflow rates and stellar mass of the galaxy is presented in the respective color in each panel. The crosshairs help visually compare the gas inflow rates for the different populations at each redshift snapshot. After infall into the host halo, gas inflow rates of both cluster and field satellites show no significant correlation with the stellar mass of the galaxy. }
    \label{fig:gas_sm_ai}
\end{figure*}
In Figure \ref{fig:gas_in} we show that at $z=2$,  the massive cluster centrals (median \logMstarMsun $= 11.2 \pm 0.2$) have higher (by $\sim0.5$ dex ) gas inflow rates compared to the field centrals and the satellite sample (median \logMstarMsun $= 10.6 \pm 0.1$). We test if the high gas inflow rate of cluster centrals are driven by their stellar mass by investigating if the gas inflow rates of a galaxy depends on their stellar mass at different redshifts and environments.


Figure \ref{fig:gas_sm_bi} shows the gas inflow rates as a function of the stellar mass of the galaxy before infall for field satellites (yellow contours, top row) and cluster satellite samples (red contours, bottom row) selected at $z=2$. The blue contours in both top and bottom rows show the comparison with gas inflow rates of the stellar mass matched field centrals selected at $z=2$. We measure the correlation between the stellar mass of the galaxy and the gas inflow rates at redshifts 6,4 and 2.6 by performing the Spearman correlation test for each sample set (summarised in Table \ref{table:corr}). 

In the early universe (at $z=6$), the Spearman correlation coefficient (r) of the gas inflow rates with the stellar mass of the field centrals, field satellites and cluster satellites are r $\geq0.7$ indicating a strong positive correlation (Table \ref{table:corr}). The strength of correlation between gas inflow rates and stellar mass of the galaxy decreases with decreasing redshift. At $z=4$, the correlation between gas inflow rates and stellar mass is moderate for field centrals and weak for the satellite sample (both field and cluster). By $z=2.6$, there is no significant correlation between the gas inflow rates with the stellar mass of the field satellites, field centrals and cluster satellites ($r<0.2$). The correlation coefficient for field centrals ($r<0.2$; Table \ref{table:corr}) at redshifts 3 to 2 also indicates no significant correlation between gas inflow rates and the stellar mass of the galaxy at $z\leq3$ (Figure \ref{fig:gas_sm_ai}).

After infall in the host halo, field and cluster satellites show no significant correlation ($r<0.2$; Table \ref{table:corr}) between the gas inflow rates and the stellar mass of the galaxy at $3\geq z\geq2$ (Figure \ref{fig:gas_sm_ai}). However, at $z=2$ a population of cluster satellites have lower gas inflow rates compared to the field centrals and field satellites. We show in sections \ref{sec:halor} and \ref{sec:overd}, that the lower gas inflow rates in the cluster satellites is correlated to their halo-centric radii and local galaxy overdensity.

We show Figure \ref{fig:gas_in} that the gas inflow rates of cluster and field centrals have similar shape, albeit the gas inflow rates in the cluster centrals is significantly higher. The lack of correlation of gas inflow rates of field centrals with their stellar mass at at redshifts 3 to 2, indicates that the gas inflow rates of the cluster centrals would not be correlated to their stellar mass at $z=2$, instead is correlated with the halo mass and the local galaxy overdensity (Figures \ref{fig:gas_halo_cen} and \ref{fig:local_over}).


\bsp	
\label{lastpage}
\end{document}